\documentclass[english,pra,a4paper,twocolumn,showpacs,preprintnumbers,superscriptaddress,amsmath,amssymb]{revtex4}

\usepackage{amsmath,amsfonts,amsthm,amssymb}
\usepackage{bbm}
\theoremstyle{plain}
\usepackage{graphicx}
\usepackage{dcolumn}
\usepackage{color}
\usepackage[colorlinks,linkcolor=blue, urlcolor=blue, anchorcolor=blue, citecolor=blue]{hyperref}
\usepackage{multirow}

\usepackage{babel}

\begin{document}
\title{Measure of the density of quantum states in information geometry and its application in the quantum multi-parameter estimation}

\author{Haijun Xing}

\affiliation{Graduate School of China Academy of Engineering Physics, Beijing, 100193, China}

\author{Libin Fu}

\email{lbfu@gscaep.ac.cn}

\affiliation{Graduate School of China Academy of Engineering Physics, Beijing, 100193, China}

\begin{abstract}
Recently, there is a growing interest in study quantum mechanics from the information geometry perspective, where a quantum state is depicted with a point in the projective Hilbert space. By taking quantum Fisher information (QFI) as the metric of projective Hilbert spaces, estimating a small parameter shift is equivalent to distinguishing neighboring quantum states along a given curve. Henceforth, information geometry plays a significant role in the single parameter estimation. However, the absence of high dimensional measures  limits its applications in studying the multi-parameter systems.   In this paper, we will discuss the physical implications of the volume element of QFI. It measures the intrinsic density of quantum states (IDQS) in projective Hilbert spaces, which is, then, a measure to define the (over) completeness relation of a class of quantum states. As an application, IDQS can be used in quantum measurement and multi-parameter estimation. Induced by the error of multi-parameter estimation, a set of estimators acquires an effective volume that is measured by the determinant of its covariance matrix. We find the density of distinguishable states (DDS) for a set of efficient estimators is thus measured by the invariant volume of the classical Fisher information, which is the classical counterpart of QFI and serves as the metric of statistical manifolds. Correspondingly, a determinant form of quantum Cram\'{e}r-Rao inequality is proposed to quantify the ability to infer the IDQS via quantum measurement. As a result, we find a gap between IDQS and maximal DDS over the measurements. The gap has tight connections with the uncertainty relationship. Exemplified with the three-level system with two parameters, we find the maximal DDS attained via the \emph{vertex measurements} (MvDDS) equals the square root of the quantum geometric tensor's determinant. It indicates the square gap between IDQS and MvDDS is proportional to the square of Berry curvature.
\end{abstract}

\maketitle

\section{Introduction}
Estimating parameters in high precision is essential for both scientific research and technical  applications. Recently, the studies of estimating multiple parameters simultaneously using quantum resources attract lots of attention \cite{Matsumoto2002,Humphreys2013,Vaneph2013,Ragy2016,Baumgrat2016,Szczykulska2016,Pezze2017,Gessner2018,Liu2020,Sidhu2019,Crowley2014,Vidrighin2014,Kok2017}.  The theory of quantum parameter estimation \cite{Helstrom1976,Holevo1982} and quantum metrology \cite{GLM2004,GLM2006,GLM2011} provides us the basic tools to estimate a single parameter via quantum measurement and the methods of enhancing the precision of parameter estimation with quantum resources. The quantum Fisher information (QFI) lies in the heart of the theory by setting the upper bounds  of a single estimator's precision via the quantum Cram\'er-Rao inequality. The single-parameter case is well-studied, and series of achievements have been made \cite{Pezze2018,Degen2017,Braun2018}, such as the high-precision magnetometry \cite{Swell2012,Ockeloen2013,Muessel2014}, atomic clocks \cite{Louchet-Chauvet2010,Leroux2010,Hosten2016,Kruse2016}, and gravitational wave detectors \cite{LIGO2011,LOGO2013} have been demonstrated in principle or realized experimentally. 

The information geometry presents us with a fundamental viewpoint to study single parameter estimation with the differential geometrical methods \cite{Rao1945,Amari1985,Amari2000,Amari2016,Wootters1981, Shapere1989,Anandan1990,Brody2001,Bengtsson2006,Braunstein1994}. 
By taking QFI as the Riemannian metric of the embedding parameter spaces, estimating a small parameter is equivalent to distinguishing neighboring quantum states along the curve given by the shift of parameter to be estimated \cite{Braunstein1994}. QFI measures the square of the density of the states distinguishable in the neighborhood of the given point (states) along the curve.  
The easiness to distinguish two states via parameter estimation is thus quantified by the statistical distance, i.e., length of the geodesic line given by QFI \cite{Wootters1981,Braunstein1994}.  
The QFI and statistical distance have tight connections with those measures widely used in quantifying the "distance" between quantum states, such as Fubini-Study metric \cite{Gibbons1992}, quantum geometric tensor \cite{Shapere1989}, quantum fidelity \cite{Braunstein1994}, and Kullback-Leibler divergence (relative entropy) \cite{Bengtsson2006}. 
As a metric, QFI also depicts the square of the speed of quantum state's "movement"  with respect to the small shift of the intrinsic or external control parameter. It is also known as fidelity susceptibility \cite{You2007,Yang2008,Gu2009,Garnerone2009,Gu2010} in those scenarios.
Therefore, in the framework of information geometry, researchers can unify topics in quantum mechanics with the parameter estimation, such as the quantum phase transition \cite{Venuti2007,Zanardi2007}, quantum non-Markovianity \cite{Lu2010},   quantum speed limit \cite{Jones2010,Zwierz2012,Taddei2013,Pires2016,Bukov2019}, optimal control \cite{Zulkowski2015,Sivak2012, Sivak2016, Tomka2016,Rotskoff2015}, and quantum algorithm \cite{Miyake2001,Cafaro2012A,Cafaro2012B}, even the thermodynamics \cite{Weinhold1975,Salamon1980,Ruppeiner1979,Ruppeiner1995,Crooks2007,Sivak2012,Zulkowski2012}. 

In general cases such as the vector magnetic field estimation \cite{Baumgrat2016}, optical imaging \cite{Humphreys2013}, and wave function detection, one simultaneously estimates more than one parameters from a given quantum state. These demands bring out the flourishing studies of the multi-parameter estimation. For the $d$-dimensional \emph{estimand} $\boldsymbol{\theta}$, i.e., parameters to be estimated, the uncertainty of the corresponding unbiased estimators is depicted by its $d\times d$ covariance matrix. 
One of the primary tasks is extracting a scalar measure out of the covariance matrix to assess the quality (precision) of those estimators and finding the saturable bounds of the measure. 
The quadratic cost function is the conventional measure widely used in nowadays studies. It is the weighted average of covariance matrix elements, by introducing a $d\times d$ non-negative definite real symmetric matrix $\boldsymbol{G}$ to weight the asymmetrical significance of the parameters \cite{Helstrom1976, Holevo1982}. The cost function is bounded by the Cram\'er-Rao-type bound \cite{Helstrom1976}, and Holevo Cram\'er-Rao bound \cite{Holevo1982,Sidhu2019}. Lots of achievements have been made with these measures \cite{Szczykulska2016,Liu2020,Matsumoto2002,Humphreys2013,Vaneph2013,Ragy2016,Gessner2018}.
 Two extreme conditions are well-studied: 1) $\boldsymbol{G}=\boldsymbol{n}\boldsymbol{n}^T$\cite{Gessner2018}. The cost function only counts the variance in a specific direction $\boldsymbol{n}$ in the parameter space and reduces to the variance of a single parameter via re-parameterization. 2) $\boldsymbol{G}$ is identity \cite{Humphreys2013,Kok2017}. The corresponding cost function is the trace of the covariance matrix. 

Estimating a set of $d$ independent parameters $\boldsymbol{\theta}$ of a given quantum state is equivalent to inferring the coordinates of a given point in $d$-dimensional parameter space. Hence the precision of the corresponding estimation highly relates to the geometrical properties of the neighborhoods of the given point $\boldsymbol{\theta}$.
However, it is hard to interpret the general cost function and its bounds as geometrical measures of the parameter space straightforwardly. The tight connections between information geometry and parameter estimation are thus loose in nowadays multi-parameter studies. It increases the difficulty in applying those results acquired in recent studies into other topics highly relates to the statistical properties of multi-parameter quantum systems. 
 
Theoretically, manifolds of  the quantum system named as the complex projective Hilbert spaces \cite{Bengtsson2006,Anandan1990,Brody2001} are intrinsically multi-dimensional. In practical studies, most of the manifolds we encountered, such as the ground states manifolds \cite{Kolodrubetz2013}, quantum phase transition \cite{Kumar2014,Banchi2014}, response theory \cite{Kolodrubetz2017,Shitara2016,Ozawa2018}, even the thermodynamics \cite{Sivak2012,Ruppeiner1979,Ruppeiner1995,Brody1995,Carollo2019} are generally multi-dimensional too. 
Well characterizing the neighborhood of a given point in the multi-dimensional manifolds is thus vital to understanding and promoting those studies. 
Hence finding a measure of multi-parameter estimation from information geometrical perspective is an essential and significant topic for the quantum information fields.

In this article, we will study the multi-parameter estimation from the information geometry perspective. 
We find, as a Riemannian metric equipped on the parameter space, QFI's volume element quantifies the \emph{intrinsic density of quantum states} (IDQS), which is a natural generalization of the "line element" in the single parameter cases. The IDQS is the measure to define the (over) completeness relation of a class of states which forms sub-manifolds of the projective Hilbert space. As its classical counterpart, the volume element of classical Fisher information presents us the \emph{density of distinguishable states} (DDS) in the statistical manifold. The DDS measures the maximal density of states that can be distinguished in a single shot of the given measurement when the quality of a set of estimators built on its results is quantified via the volume occupied by their "error ball". 
The IDQS bounds the DDS via the quantum Cram\'er-Rao inequality in determinant form.  Different from the single parameter cases, this bounds is not always attainable. A gap between the IDQS and the maximal DDS achieved via quantum measurement is found. We will study the three-level system as an example, which is the minimal system to study the gap. As a result, a tight connection between the gap and the Berry curvature is found.

This article is organized as follows. 
In Sec.~\ref{review}, we review the single parameter estimation from the information geometry perspective. In Sec.~\ref{DQS}, the DDS and IDQS are introduced. In Sec.~\ref{QCRID}, the ability to inter the IDQS with the quantum measurements is studied via quantum Cram\'er-Rao inequality in the determinant form. As a result, a gap between the maximal DDS and the IDQS is found. In Sec.~\ref{3LS}, the three-level system is proposed to study the gap, and the tight connection between the gap and Berry curvature will be shown. At last, we summarize this article.  

\section{\label{review}Review of quantum geometric tensor and single parameter estimation}
In quantum mechanics, one usually terms the state space of an $(n+1)$-level system as the $(n+1)$-dimensional Hilbert spaces. 
However, an additional equivalence $|\psi\rangle\sim c|\psi\rangle$, with $c\in\mathbb{C}\backslash\{0\}$, is assumed implicitly. It depicts the demands of normalization and the physical insight that two states only different in the global phases are indistinguishable. Under this equivalence, the actual state space we handle is the so-called projective Hilbert spaces $\mathbb{C}\boldsymbol{\mathrm{P}}^n$ or its sub-manifold generally \cite{Bengtsson2006,Anandan1990,Brody2001}.
Therefore, one usually parameterizes the quantum states with a model $\mathcal{M}=\{\left.|\psi(\boldsymbol{\theta}\rangle\langle\psi(\boldsymbol{\theta})|\right|\boldsymbol{\theta}\in\Theta\}$,  which gives a real coordinate system $\boldsymbol{\theta}=\{\theta^1,\theta^2,\dots,\theta^d\}^T$, with $d\leqslant 2n$, to (the sub-manifold of) $\mathbb{C}\boldsymbol{\mathrm{P}}^n$ effectively.
The movement along the "radial direction" of state $|\psi(\boldsymbol{\theta})\rangle$ is null under this equivalence. 
Based on that, the intrinsic derivative is given by 
\begin{equation}
\hat{\nabla}_{\mu}|\psi\rangle \equiv\left(\hat{\mathbbm{1}}-|\psi\rangle\langle\psi|\right)\hat{\partial}_{\mu}|\psi\rangle,
\end{equation}
with $\hat{\partial}_{\mu}\equiv\partial/\partial\theta^{\mu}$, and $|\psi\rangle\equiv|\psi(\boldsymbol{\theta})\rangle$ for succinctness.   The normalization $\langle\psi |\psi\rangle =1$ is assumed. The derivative is orthogonal to the state $|\psi\rangle$ with $\langle\psi|\hat{\nabla}_\mu |\psi\rangle=0$. In this form, the quantum geometric tensor $\boldsymbol{\mathcal{Q}}$ is defined by \cite{Shapere1989}
\begin{equation}
\mathcal{Q}_{\mu\nu}\equiv \langle\psi|\hat{\overleftarrow{\nabla}}_{\mu}\hat{\nabla}_{\nu}|\psi\rangle
= g^F_{\mu\nu}+\mathrm{i}\sigma_{\mu\nu},
\end{equation}
where the antisymmetric part $\sigma_{\mu\nu}\equiv\mathrm{i}\mathcal{Q}_{[\nu\mu]}=-\mathcal{B}_{\mu\nu}/2$ is proportional to the Berry curvature $\mathcal{B}_{\mu\nu}$;  the symmetric part  $g^F_{\mu\nu}\equiv \mathcal{Q}_{\{\mu\nu\}}$ severs as the Riemannian metric of the projective Hilbert spaces  when $\mathbb{C}\boldsymbol{\mathrm{P}}^n$ is treated as a $2n$-dimensional real manifold. We denote $\boldsymbol{g}^F$ as the \emph{quantum Fisher metric} (QFM) in this article, for $\boldsymbol{g}^F$ is a quarter of the  quantum Fisher information (QFI) $\boldsymbol{\mathcal{F}}$. 
The QFM defines the
statistical distance with \cite{Wootters1981,Braunstein1994}
\begin{equation}
ds^{2}\equiv g^F_{\mu\nu}\dot{\theta}^{\mu}\dot{\theta}^{\nu}dt^2=g^F_{tt}dt^2,
\end{equation}
where $\dot{\theta}^\mu=d\theta^\mu/dt$ is the derivative along the curve $\boldsymbol{\theta}(t)$, and the Einstein summation convention is assumed. 
The length of a curve acquired by integrating the element $ds$ depicts the maximal number of states distinguishable along the curve. The corresponding distance measures the easiness of distinguishing the quantum states via quantum parameter estimation.
\begin{figure}
\centering
\includegraphics[width=8.5cm]{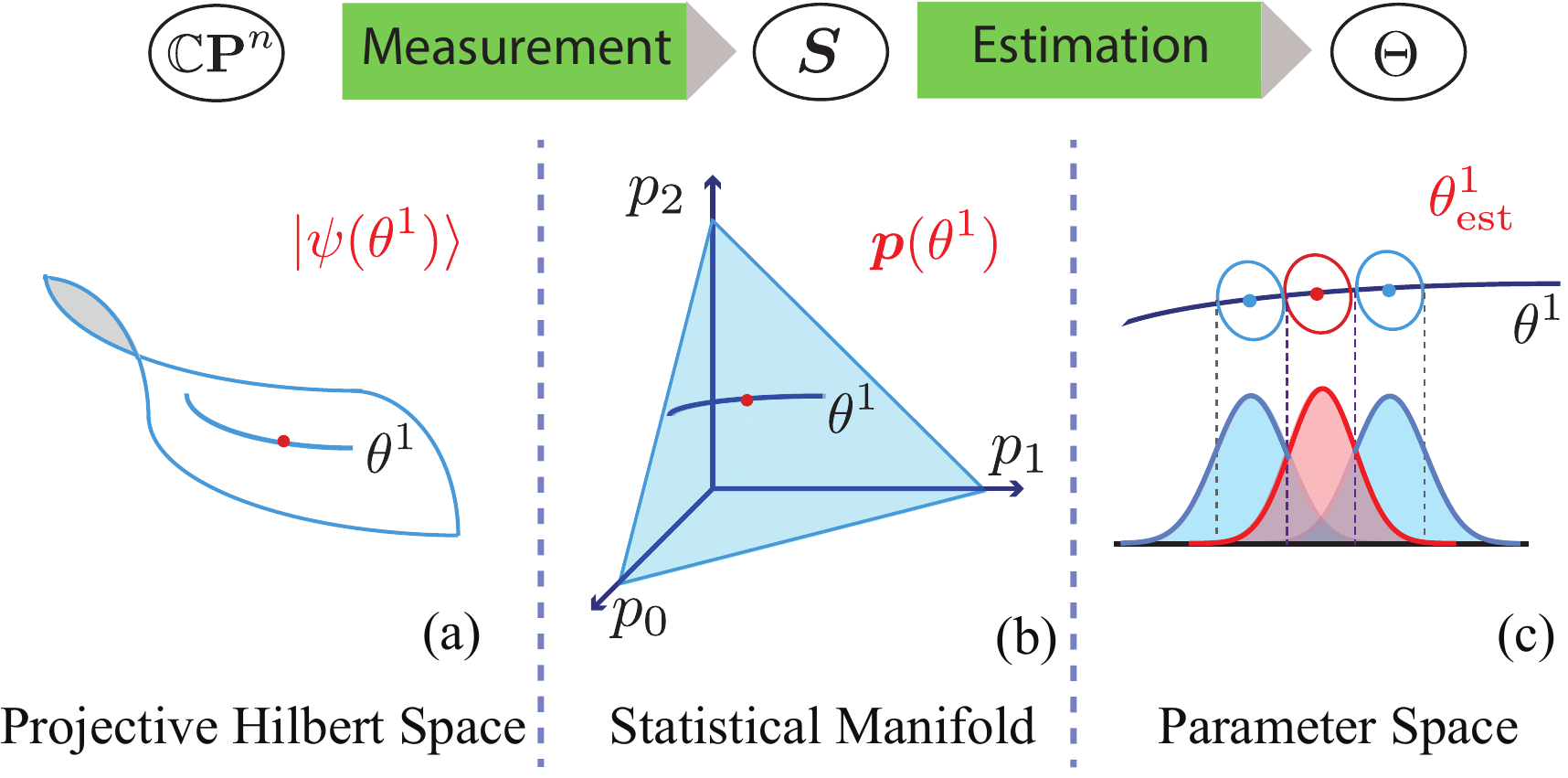}
\caption{(color online). Quantum single parameter estimation (exemplified with the three-level system). 
(a) The state of the system locates in the projective Hilbert space $\mathbb{C} \boldsymbol{\mathrm{P}}^2$. Only one real parameter $\theta^1$ is assumed unknown and needs estimation. The state $|\psi(\theta^1)\rangle$ draws a curve in $\mathbb{C} \boldsymbol{\mathrm{P}}^2$ via variation of $\theta^1$.  The statistical length of the curve is defined with the element $ds^2=g^F_{11}(d\theta^1)^2$.   
(b) Via a ternary-outcome projective measurement $\{\hat{E}_i|i=0,1,2\}$, the state $|\psi(\theta^1)\rangle$ reduces to a classical distribution $\boldsymbol{p}(\theta^1)=(p_0,p_1,p_2)$ with $p_i=\langle\psi(\theta^1) |\hat{E}_i|\psi{(\theta^1)} \rangle$. $\boldsymbol{p}(\theta^1)$ locates in a curve in the statistical manifold, which is a two-simplex. The density of classical distribution along the curve is measured  by $ds/d\theta^1=\sqrt{g^I_{11}}$. The maximum of this density is $\sqrt{g^F_{11}}$, which can be reached via the optimal measurement.  
(c) One estimates the state, i.e., the parameter $\theta^1$, with the sample acquired from a sequence of identical measurements. The width of the ``error ball'' of $\theta^1_{\rm est}$ along the curve in the parameter space $\Theta$ is measured by $2\delta\theta^1_{\rm est}$. 
Two distributions can be reliably distinguished when their error balls have no overlaps. If the estimation is efficient, the density of states distinguishable in a single shot of the given measurement is maximal, which equals $\sqrt{g^I_{11}}$. }
\label{Fig1}
\end{figure}

In the process of parameter estimation as shown by Fig.~(\ref{Fig1}) and Fig.~(\ref{Fig2}), the state $|\psi(\boldsymbol{\theta})\rangle$ is inferred  via a set of positive operator-valued measurement (POVM) $\hat{\boldsymbol{E}}=\{\hat{E}_{i}\}$ with $\sum_i \hat{E}_i=\hat{\mathbbm{1}}$. The result $i$ is acquired with probability $p_{i}=\langle\psi(\boldsymbol{\theta})|\hat{E}_{i}|\psi(\boldsymbol{\theta})\rangle$.
Mathematically, the measurement reduces the projective Hilbert spaces of states $|\psi(\boldsymbol{\theta})\rangle$ to a statistical manifold of classical distribution $\boldsymbol{p}=(p_{0},p_{1},\dots p_{n})$. Corresponding to the QFM, the metric  of the statistical manifold is the \emph{Fisher-Rao metric} (FRM) $\boldsymbol{g}^I$ with the elements
\begin{equation}
g^I_{\mu\nu}=\sum_i \partial_{\mu} \sqrt {p_{i}}\partial_{\nu}\sqrt{p_{i}}.
\end{equation}
FRM is a quarter of the classical Fisher information (CFI) $\boldsymbol{\mathcal{I}}$.
 One builds estimators $\boldsymbol{\theta}_{\mathrm{est}}$ of the parameters $\boldsymbol{\theta}$ with the sample acquired  by the measurement results after $m$ repetition. The precision of the estimators $\boldsymbol{\theta}_{\rm est}$ is measured by $ \boldsymbol{\Sigma}^{-1}$, the inverse of its covariance matrix $\boldsymbol{\Sigma}$ with $\Sigma^{\mu\nu}=\mathrm{Cov}(\theta_{\mathrm{est}}^\mu,\theta_{\mathrm{est}}^\nu)$.  CFI sets the upper bound of the precision; and CFI itself  is upper bounded by QFI via the quantum Cram\'er-Rao inequality (QCRI) \cite{Helstrom1976,Holevo1982}
\begin{align}
m\boldsymbol{\mathcal{F}}\geqslant m\boldsymbol{\mathcal{I}}\geqslant \boldsymbol{\Sigma}^{-1},\label{QCR1}
\end{align}
it indicates $m\mathcal{\mathcal{F}_{\mu\mu}}\geqslant m\mathcal{I}_{\mu\mu}\geqslant(\boldsymbol{\Sigma}^{-1})_{\mu\mu}=1/\delta^2\theta^\mu_{\rm est}$, when only one of the parameters, e.g., $\theta^\mu$ as shown by Fig.~(\ref{Fig1}),  needs estimation. The saturation of the ultimate precision $m\mathcal{F}_{\mu\mu}$ needs  optimizing both of the estimation and measurement: 
the last equality  is reached by maximally likelihood estimation, the first equality is reached by the optimal measurement $\hat{\boldsymbol{E}}$ satisfying \cite{Braunstein1994}
\begin{equation}
|\psi\rangle\langle\psi|(\lambda_{\mu}-\hat{L}_{\mu})\hat{E_{i}}^{1/2}=0,\forall i,\label{om}
\end{equation}
with $\lambda_{\mu}\in \mathbb{R}$. $\hat{L}_\mu$ is the symmetric logarithmic derivative (SLD) defined by $\hat{\nabla}_\mu\hat{\rho}\equiv(\hat{L}_\mu \hat{\rho} +\hat{\rho} \hat{L}_\mu)/2$ with $\hat{\rho}=|\psi\rangle\langle \psi|$. 
We mention that the QCRI Eq.~(\ref{QCR1}) is still valid, when $\hat{\rho}$ is a general mixed state  \cite{Helstrom1976,Holevo1982,Braunstein1994}.

In multi-parameter cases, one needs simultaneously estimate a set of parameters from the given state $|\psi(\boldsymbol{\theta})\rangle$. The QCRI Eq.~(\ref{QCR1}) is still valid. However, to quantify the quality of measurement and estimation, one should extract a scalar index out of each matrix of Eq.~(\ref{QCR1}).  The index in the traditional framework is the weighted average of the covariance matrix elements by introducing a real symmetric positive cost matrix $\boldsymbol{G}$.  It brings us the inequality
\begin{equation}
m\mathrm{tr}(\boldsymbol{G}\boldsymbol{\mathcal{F}}^{-1})\leqslant m\mathrm{tr}(\boldsymbol{G}\boldsymbol{\mathcal{I}}^{-1}) \leqslant \mathrm{tr}(\boldsymbol{G}\boldsymbol{\Sigma}),
\end{equation}
without loss of the generality,  all of the components of $\boldsymbol{\theta}$ is assumed unknown, and independent of each other. It indicates that both $\boldsymbol{\mathcal{F}}$ and $\boldsymbol{\mathcal{I}}$ are full rank and completely positive. The necessary and sufficient condition of saturating the bound is  $\langle \psi(\boldsymbol{\theta})| [\hat{L}_\mu,\hat{L}_\nu]|\psi(\boldsymbol{\theta}) \rangle=0$, $\forall \mu,\nu$. It is the \emph{commutation condition}  proved by Matsumoto \cite{Matsumoto2002}.
Specifically, when $\boldsymbol{G}$ is taken as the identity,  the corresponding index is the trace of the covariance matrix, which is widely used in recent studies \cite{Matsumoto2002,Humphreys2013,Vaneph2013,Ragy2016,Gessner2018}. 
Furthermore, we mention that a tighter multi-parameter bound of $\mathrm{tr}(\boldsymbol{G\Sigma})$ named as the Holevo bound \cite{Holevo1982} is widely used too. 

\section{\label{DQS} Density of states}
The Riemannian geometry provides us a standard method to quantify the invariant volume and the corresponding density of a Riemannian manifold. If $\boldsymbol{g}$ serves as the metric of a Riemannian manifold $(M,\boldsymbol{g})$ with coordinates $\Theta$, $dV=\sqrt{|\boldsymbol{g}|}d\Theta$ defines the invariant volume element of the manifold $M$, where $|\boldsymbol{g}|$ denotes the determinant of $\boldsymbol{g}$. The element $dV$ is invariant under the change of coordinates. It indicates $\sqrt{|\boldsymbol{g}|}=dV/d\Theta$ measures an intrinsic density of the manifold. Hence, in the framework of information geometry, one can formally define a measure of the density of states in a statistical manifolds (projective Hilbert spaces) with $\sqrt{|\boldsymbol{g}^I|}$ ($\sqrt{|\boldsymbol{g}^F|}$).
The two densities have ample physical implications. As we will show below, they  naturally emergent from the basic theory of multi-parameter estimation as the bounds of precision measure.

\subsection{Volume of estimators and density of distinguishable states} 
In multi-parameter estimation, researchers simultaneously estimate a set of $d$ independent parameters, i.e., the \textit{estimand} $\boldsymbol{\theta}$ from the distribution $\boldsymbol{p}(\boldsymbol{\theta})=(p_0,p_1,\dots,p_{n})$, with $p_i=\langle\psi(\boldsymbol{\theta})|\hat{E}_i|\psi(\boldsymbol{\theta})\rangle$.
After $m$ repetitions of trails, one acquires a sample with $m$ measurement results, in which the outcome $i$ occurs with frequency $\xi_i$.
According to the central limit theorem, distribution of the frequency $\boldsymbol{\xi}=(\xi_0,\xi_1,\dots,\xi_{n})$ converges to a Gaussian distribution 
\begin{equation}
\rho(\boldsymbol{\xi}|\boldsymbol{\theta})\propto\exp\left[-\frac{m}{2}\sum_i\frac{(p_i-\xi_i)^2}{p_i}\right],
\end{equation}
with $m\rightarrow\infty$. This distribution is highly localized in the neighborhood of the true value $\boldsymbol{p}(\boldsymbol{\theta})$. It is natural to conjecture that there exist a set of unbiased estimators $\bar{\boldsymbol{\theta}}_{\rm{est}}(\boldsymbol{\xi})$ such that with the repetition  $m\rightarrow\infty$, the distribution $\rho(\bar{\boldsymbol{\theta}}_{\rm{est}}|\boldsymbol{\theta})$ is asymptotic to
\begin{equation}
\rho(\bar{\boldsymbol{\theta}}_{\rm est}|\boldsymbol{\theta}) \propto \exp{\left[-\frac{4m}{2}(\bar{\boldsymbol{\theta}}_{\rm est}-\boldsymbol{\theta})^T\boldsymbol{g}^I(\bar{\boldsymbol{\theta}}_{\rm est}-\boldsymbol{\theta})\right]}\label{AE}
\end{equation}
in the neighborhood of $\boldsymbol{\theta}$, where the linear approximation $p_i-\xi_i\approx \partial_\mu p_i (\theta^\mu-\bar{\theta}^\mu_{\rm{est}})$ is valid.  
The validity of this conjecture in the whole parameter space relates to the topics of asymptotic normality of estimation, where $\bar{\boldsymbol{\theta}}_{\rm est}$ with distribution Eq.~(\ref{AE}) are called as the asymptotically efficient estimators. 
Roots of likelihood equations and maximum likelihood estimation are proved efficient asymptotically under the regularity conditions which indicate \cite{Lehmann1998,JunShao2003}:
\begin{description}
 \item[{a1}] The estimators $\bar{\boldsymbol{\theta}}_{\rm est}(\boldsymbol{\xi})$ are well-defined as the single-valued functions of $\boldsymbol{\xi}$. 
 \item[{a2}] The FRM $\boldsymbol{g}^I$ is positive definite for all $\boldsymbol{\theta}\in\Theta$, and the elements $\boldsymbol{g}^I_{\mu\nu}$ are finite.
 \item[{a3}] The third derivatives $\partial_\mu\partial_\nu\partial_\gamma\log[\rho(\boldsymbol{\xi}|\boldsymbol{\theta})]$ exist and are  bounded for all $\mu,\nu, \gamma$, and $\boldsymbol{\theta}\in\Theta$. 
 \end{description}
 Theoretically, one can narrow $\Theta$ and the range of estimators to arbitrary small open subset containing $\boldsymbol{\theta}$ with sufficient prior information. The regularity conditions are thus satisfied by most of the statistical models in quantum metrology. Hence, in this article, we assume the asymptotically efficient estimators $\bar{\boldsymbol{\theta}}_{\rm est}$ are always exist.

Two Gaussian distributions can be  reliably discriminated when their overlap less than a specific value, as shown by Fig.~\ref{Fig1} (c). The distribution $\rho(\bar{\boldsymbol{\theta}}_{\rm est}|\boldsymbol{\theta})$ thus acquires an effective width along a given curve  $\boldsymbol{\theta}(t)$. Then a finite number of states are distinguishable on a segment of the curve. It is the core ingredient of  the statistical distance \cite{Wootters1981}. For general unbiased estimators $\boldsymbol{\theta}_{\rm est}$, the FRM $\boldsymbol{g}^I$ still bounds inverse of theirs' covariance matrix $\boldsymbol{\Sigma}$ as shown by the QCRI Eq.~(\ref{QCR1}). It indicates the distribution of $\boldsymbol{\theta}_{\rm est}$  still highly localized. The variance $\boldsymbol{\Sigma}$ is still a qualified measure of $\boldsymbol{\theta}_{\rm est}$'s uncertainty, with the repetition $m\rightarrow\infty$.
 In consistence with Wootters \cite{Wootters1981}, we take the width of $\rho(\boldsymbol{\theta}_{\rm est}|\boldsymbol{\theta})$ along the curve $\boldsymbol{\theta}(t)$ as $2\delta t$, with the variance $\delta t\equiv \mathrm{tr}[\dot{\boldsymbol{\theta}}\dot{\boldsymbol{\theta}}^T\boldsymbol{\Sigma}]^{1/2}$. 

\begin{figure}[ptb]
\centering
\includegraphics[width=8.5cm]{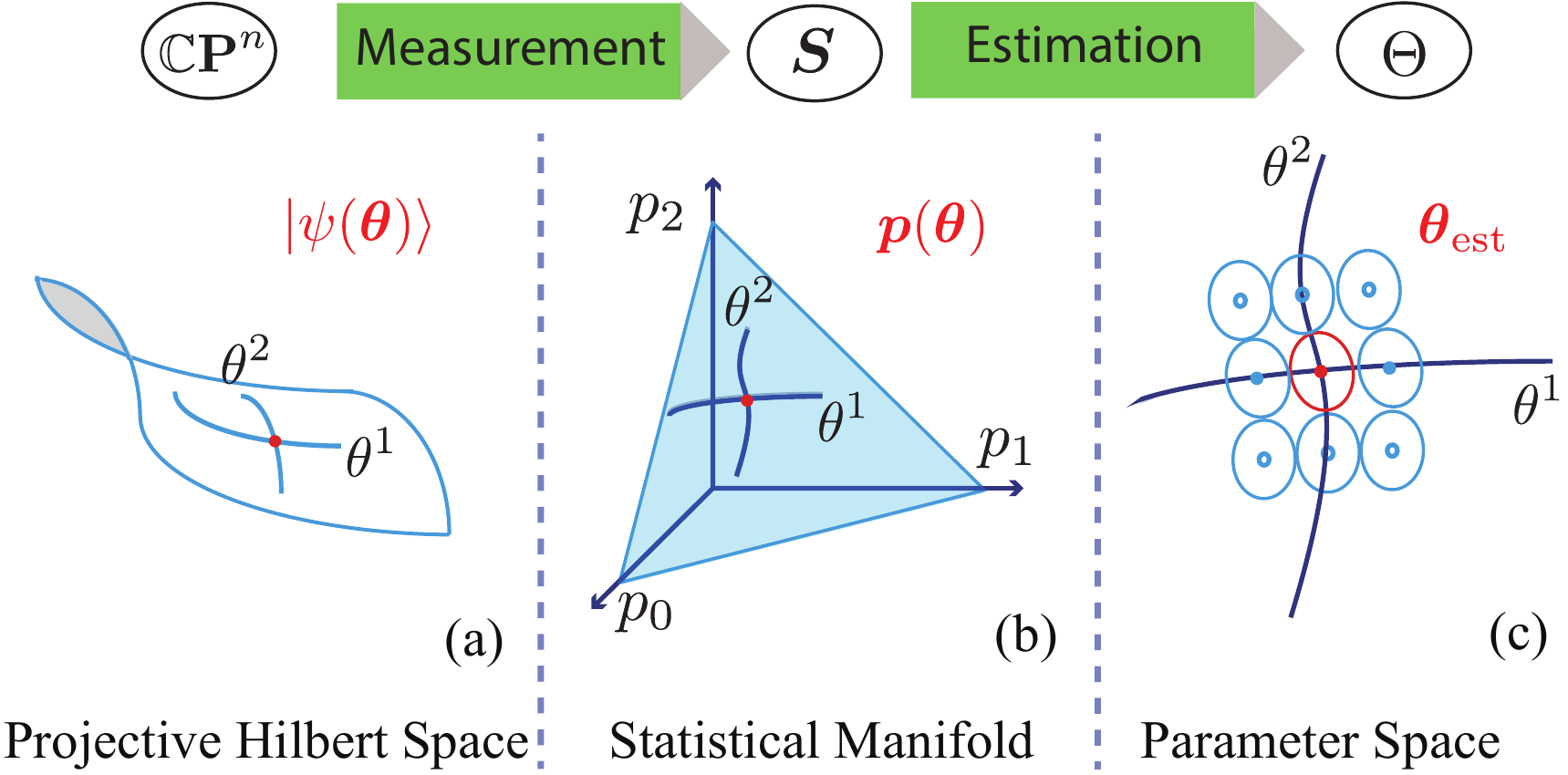}
\caption{(color online).  Quantum multi-parameter estimation (exemplified with the three-level system). 
(a) State $|\psi(\boldsymbol{\theta})\rangle$ of the system locates in the projective Hilbert spaces $\mathbb{C} \boldsymbol{\mathrm{P}}^2$. We focus on its two-dimensional sub-manifolds, where only two real parameters $\boldsymbol{\theta}=\{\theta^1,\theta^2\}$ are assumed unknown and need estimation.  The IDQS in this space is measured by $\sqrt{|\boldsymbol{g}^F|}$.   
(b) Via a ternary-outcome projective measurement $\{\hat{E}_i|i=0,1,2\}$, the state $|\psi(\boldsymbol{\theta})\rangle$ reduces to the classical distribution $\boldsymbol{p}(\boldsymbol{\theta})=(p_0,p_1,p_2)$ with $p_i=\langle\psi(\boldsymbol{\theta}) |\hat{E}_i|\psi{(\boldsymbol{\theta})} \rangle$. $\boldsymbol{p}(\boldsymbol{\theta})$ locates in a statistical manifold, which is a two-simplex. $\sqrt{|\boldsymbol{g}^I|}$ measures the DDS in this simplex.  
(c) One estimates the state, i.e., the parameters $\boldsymbol{\theta}$, with the sample acquired from a sequence of identical measurements. The volume occupied by the ``error ball'' of $\boldsymbol{\theta}_{\rm est}$ in the parameter space $\Theta$ is measured by $\sqrt{|4\boldsymbol{\Sigma}|}$. 
Two distributions can be reliably distinguished when their error balls have no overlaps. If the estimation is efficient, the density of states distinguishable in a single shot of the given measurement is maximal, which equals the DDS $\sqrt{|\boldsymbol{g}^I|}$. }
\label{Fig2}
\end{figure}

In the multi-parameter cases as shown in Fig.~\ref{Fig2} (c), all of the $d$ components of $\boldsymbol{\theta}$ are assumed unknown.
The distribution $\rho(\boldsymbol{\theta}_{\rm{est}}|\boldsymbol{\theta})$ expandes in all directions, hence endowed an effective volume in the $d-$dimensional parameter space. For the covariance matrix is a primary measure of the estimators' uncertainty,
we take $V_E(\boldsymbol{\theta_{\rm est}})\equiv\sqrt{|4\boldsymbol{\Sigma}|}$ as a measure of the volume of the distribution $\rho(\boldsymbol{\theta}_{\rm est}|\boldsymbol{\theta})$, henceforth $\boldsymbol{p}(\boldsymbol{\theta})$. The number of states distinguishable in the neighborhood $d\Theta$ of point $\boldsymbol{\theta}$  is thus  measured by $d\Theta/V_E(\boldsymbol{\theta}_{\rm est})$. 
It is vivid in the diagonal coordinates $\boldsymbol{\zeta}$ of the covariance matrix $\boldsymbol{\Sigma}$, where estimators' volume equals $\Pi_\mu2\delta\zeta^\mu$. $\Pi_\mu n^\mu$ states can be distinguished reliably in a volume element $\Pi_\mu d\zeta^\mu$ totally, with $n^\mu=d\zeta^\mu/2\delta\zeta^\mu$ states distinguishable out of the increment $d\zeta^\mu$.

Based on the above discussions, we define $\sqrt{|\boldsymbol{g}^I|}$ as the local \emph{density of distinguishable states} (DDS) in the neighborhood of point $\boldsymbol{\theta}$. 
It is a natural generalization of the statistical distance. 
The DDS measures the maximal density of estimators $\boldsymbol{\theta}$, i.e., quantum states $|\psi(\boldsymbol{\theta})\rangle$ distinguishable  in a single shot measurement with
\begin{equation}
\sqrt{m^d|\boldsymbol{g}^I|}\geqslant 1/V_E(\boldsymbol{\theta}_{\mathrm{est}}), \label{VE} 
\end{equation}
where the constant $\sqrt{m^d}$ denotes the enhancement  of repetitions,
and  the equality is reached by efficient estimators $\bar{\boldsymbol{\theta}}$ with $1/(4\boldsymbol{\Sigma})=m\boldsymbol{g}^I$.
The proof will be given with the QCRI in Eq.~(\ref{DCR}).
Furthermore, we mention that $\sqrt{|\boldsymbol{\mathcal{I}}|}d\Theta$ is also well-known as the Jeffreys prior  \cite{Jeffreys1946,Jeffreys1948,Jaynes1968,Jaynes2003} in Bayesian estimation. It is the non-informative prior distribution in the parameter space $\Theta$.

\subsection{Intrinsic density of quantum states}
The same as the FRM, the QFM $\boldsymbol{g}^F$ serves as the metric of the projective Hilbert spaces $\mathbb{C}\boldsymbol{\mathrm{P}}^n$, and $\sqrt{|\boldsymbol{g}^F|}$ measures the \emph{intrinsic density of quantum states}  (IDQS) in $\mathbb{C}\boldsymbol{\mathrm{P}}^n$ with 
\begin{equation}
dV_q/d\Theta=\sqrt{|\boldsymbol{g}^F|},
\end{equation}
 where $dV_q$ denotes the invariant volume element of $\mathbb{C}\boldsymbol{\mathrm{P}}^n$.
The form of IDQS is invariant under re-parametrization, and its value is invariant under SU(N) rotation in Hilbert spaces.
The IDQS depicts the ``uniformity" of $\mathbb{C}\boldsymbol{\mathrm{P}}^n$. For  each point $\boldsymbol{\theta}$ in the parameter space $\Theta$,  there exists a projector $|\psi(\boldsymbol{\theta})\rangle\langle\psi(\boldsymbol{\theta}) |$  illustrates the projection to states $|\psi(\boldsymbol{\theta})\rangle$ in projective Hilbert spaces. Together with the IDQS serving as the intrinsic measure, one can  define a projector to the projective Hilbert spaces with 
\begin{equation}
\hat{\mathbbm{1}}\propto\int d\Theta\sqrt{|\boldsymbol{g}^F(\boldsymbol{\theta})|}|\psi(\boldsymbol{\theta})\rangle\langle\psi(\boldsymbol{\theta})|,\label{CR1}
\end{equation}
if $\Theta$ and $\mathbb{C}\boldsymbol{\mathrm{P}}^n$ are isomorphic. It is indeed the completeness relation, or decomposition of $\mathbbm{1}$ of the projective Hilbert spaces \cite{Bengtsson2006}. A stretch of the proof of Eq.~(\ref{CR1}) is given in App.~\ref{ProofCompleteness}.

In practical studies, one often deals with a class of states, such as coherent states and spin squeezed states, which composes a sub-manifold of the projective Hilbert spaces. The density of quantum states is inherited from $\mathbb{C}\boldsymbol{\mathrm P}^n$ together with the induced metric. Hence if a class of parameterized states is complete (overcomplete), one may calculate the completeness   relation with Eq.~(\ref{CR1}) by integrating over the parameter space $\Theta$.  The examples of coherent states and squeezed states are given in App.~\ref{App:ECR}.
It is a new method that can significantly decrease the complexity of calculating the completeness. 

\section{\label{QCRID}Quantum Cram\'er-Rao Inequality in Determinant Form}
Via quantifying the density of states attained in a given measurement, DDS serves as a measure of the measurement's quality in multi-parameter estimation. Eq.~(\ref{QCR1}) indicates that IDQS upper bounds the DDS over the sets of POVM for a given quantum state.  Specifically, we generalize the QCRI Eq.~(\ref{QCR1}) to the determinant form
\begin{equation}
\sqrt{m^d|\boldsymbol{g}^F|}\geqslant\sqrt{m^d|\boldsymbol{g}^I|}\geqslant1/\sqrt{|4\boldsymbol{\Sigma}|},\label{DCR}
\end{equation}
where the equalities are reached if the matrices on two sides of the respective inequality are equal.

\begin{proof}
We begin with two arbitrary positive definite real symmetric matrices $\boldsymbol{A}$ and $\boldsymbol{B}$, which satisfy the matrix inequality $\boldsymbol{A}\geqslant \boldsymbol{B}$, i.e., 
\begin{equation}
\boldsymbol{A}-\boldsymbol{B}\geqslant\boldsymbol{0}.\label{MathI}
\end{equation}
One can diagonalize the difference matrix with a unitary  matrix $\boldsymbol{U}$. Denoting $\boldsymbol{UMU}^{-1}=\boldsymbol{M}',\boldsymbol{M}=\boldsymbol{A},\boldsymbol{B}$,
we have
\begin{equation}
\boldsymbol{C}\equiv \boldsymbol{A}'-\boldsymbol{B}'=\mathrm{diag}\left[\lambda_{1},\lambda_{2},\dots,\lambda_{d}\right],
\end{equation}
with the eigenvalue $\lambda_{i}\geqslant0,$ for all ${i}$. For $\boldsymbol{B}'$ is positive definite, we have $|[\boldsymbol{B}']^{ij\dots}|>0$, where $[\boldsymbol{B}^{'}]^{ij\dots}$ is the algebraic complement of $\{B'_{ii},B'_{jj},\dots\}$.  Therefore, we have the determinant
\begin{eqnarray}
|\boldsymbol{A}'|&= & |\boldsymbol{B}'+\boldsymbol{C}|\nonumber \\
&= & |\boldsymbol{B}'|+\lambda_{i}|[\boldsymbol{B}']^{i}|+\lambda_{i}\lambda_{j}|[\boldsymbol{B}']^{ij}|+...+\Pi_{i}\lambda_{i}\nonumber \\
&\geqslant & |\boldsymbol{B}'|,\label{SIP}
\end{eqnarray}
which indicates
\begin{equation}
|\boldsymbol{A}|\geqslant |\boldsymbol{B}|=1/|\boldsymbol{B}^{-1}|.
\end{equation}
The equality holds iff $\lambda_{i}=0$, $\forall i$, i.e., $\boldsymbol{A}=\boldsymbol{B}$.
By setting $\boldsymbol{A}=\boldsymbol{g}^F$, and $\boldsymbol{B}=\boldsymbol{g}^I$ ($\boldsymbol{A}=m\boldsymbol{g}^I$, and $\boldsymbol{B}=(4\boldsymbol{\Sigma})^{-1}$), the first (second) inequality in Eq.~(\ref{DCR}) is thus proved.
\end{proof}

Physically, Eq.~(\ref{DCR}) depicts the ability to infer the density of quantum states via the quantum measurement.  
For the single parameter cases, the upper bound defined by the QFM is exact, as shown by Eq.~(\ref{QCR1}). It indicates one can infer the QFM element $f_{\mu\mu}$ via the quantum measurement without the loss of the distinguishability of the quantum states.  However, the situation for the IDQS in multi-parameter cases is more complicated, as will be shown below.

\subsection{The gap between maximal DDS and IDQS}
 
To attain the first equality in Eq.~(\ref{DCR}) for a given state, one should perform a measurement that is simultaneously optimal for all of the components of parameters $\boldsymbol{\eta}$, the diagonal coordinates of QFM $\boldsymbol{g}^F$. However, such a  measurement does not always exist for a general state $|\psi(\boldsymbol{\eta})\rangle$.  The critical point  is the optimal measurement of each specific component $\eta^\mu$ may non-commute with each other. 
The attainability  condition is consistent with the well-known compatibility condition  \citep{Matsumoto2002, Pezze2017,Ragy2016}, which states the optimal measurement corresponding to two parameters $\theta^\mu$ and $\theta^\nu$ are compatible only if the Berry curvatures $\mathcal{B}_{\mu\nu}\equiv\langle\psi\boldsymbol{(\eta)}|[\hat{L}_\mu,\hat{L}_\nu]|\psi(\boldsymbol{\eta})\rangle/4$ vanishes on state $|\psi(\boldsymbol{\eta})\rangle$.
It also indicates the IDQS is only attainable for states $|\psi(\boldsymbol{\eta})\rangle$ with the vanishing Berry curvature $\mathcal{B_{\mu\nu}}$, $\forall\mu,\nu$. 

For states with non-zero Berry curvature, the maximal DDS attained over the measurements is smaller than the corresponding IDQS. 
A gap between maximal DDS and IDQS  is induced by the  incompatibility of the optimal measurement of those parameters. 
A naive conjecture is the gap depending on the Berry curvature.
Next, we will show it is valid for \emph{vertex measurements} on the  three-level system with two parameters.

\section{\label{3LS}Three-level systems}

To study the gap between maximal DDS and IDQS, we need at least two independent parameters.
It indicates the minimal quantum system is three-level, which can support ternary-outcome projective measurements and induces the classical distribution locating in a two-simplex, as shown in Fig.~\ref{Fig2}. 
The projective Hilbert space $\mathbb{C}\boldsymbol{\mathrm{P}}^2$ of the three-level system is four-dimensional in real coordinates, i.e.,  the pure state of these systems has four independent parameters. We will study its two-dimensional sub-manifolds by fixing the other two parameters of the four. 

\subsection{Vertex measurements}
Even in the single parameter cases, finding a measurement scheme optimal for arbitrary given states  is complicated. However, if sufficient prior information is provided, one can apply an asymptotically optimal measurement scheme: projective measurement $\hat{\boldsymbol{E}}^v(\theta^\mu)$ with the state $|\psi(\theta^\mu)\rangle\langle\psi(\theta^\mu)|\in \hat{\boldsymbol{E}}^v(\theta^\mu)$ is asymptotically optimal for the given state $|\psi(\theta^\mu+\delta\theta^\mu)\rangle$ with the mismatch $\delta\theta^\mu$ approaches zero.
In multi-parameter estimation, Humphreys \emph{et al.} \cite{Humphreys2013} and Pezz\'e \textit{et al.} \cite{Pezze2017} prove that $\hat{\boldsymbol{E}}^v(\boldsymbol{\theta})\equiv\{|\Upsilon_{i}\rangle\langle \Upsilon_{i}|\}$ with $|\Upsilon_{0}\rangle\sim|\psi(\boldsymbol{\theta})\rangle$ is also asymptotically optimal for state $|\psi(\boldsymbol{\theta}+\delta\boldsymbol{\theta})\rangle$ with zero Berry curvature \cite{Liu2020}.  For the distribution of its measurement results $\boldsymbol{p}$ locates in the neighborhood of vertex $(1,0,0,\dots)$ of the simplex, we denote $\hat{\boldsymbol{E}}^v(\boldsymbol{\theta})$ as the \emph{vertex measurement} for state $|\psi(\boldsymbol{\theta}+\delta \boldsymbol{\theta})\rangle$ in this article for convenience.  The most straightforward vertex measurement is a binary measurement $\{|\psi(\boldsymbol{\theta})\rangle\langle\psi(\boldsymbol{\theta})|, \hat{\mathbbm{1}}-|\psi(\boldsymbol{\theta})\rangle\langle\psi(\boldsymbol{\theta}) |\}$ which only recognizes $|\psi(\boldsymbol{\theta})\rangle$ and its complementary spaces. For the parameter $\theta^\mu$,  there exists an informative vertex measurement
\begin{equation}
\hat{\boldsymbol{E}}^\mu(\boldsymbol{\theta})=\left\{|\psi(\boldsymbol{\theta})\rangle\langle\psi(\boldsymbol{\theta})|,|\nabla_{\mu}\psi\rangle\langle\nabla_{\mu}\psi|, \hat{E}_2, \dots \right\},
\end{equation}
where $|\nabla_{\mu}\psi\rangle\equiv \hat{\nabla}_{\mu}|\psi(\boldsymbol{\theta})\rangle/\lambda_{\mu}$ 
denotes the ``direction of speed'' of the state $|\psi(\boldsymbol{\theta})\rangle$'s movement in projective Hilbert spaces induced by the shift of parameter $\theta^\mu$, the norm $\lambda_{\mu}=(g^F_{\mu\mu})^{1/2}$
is the corresponding ``velocity''. Furthermore, $\hat{E}_2$, $\hat{E}_3$, $\dots$ have no contribution to the estimating of $\delta\theta^\mu$ in single parameter estimation, for the movement are confined in the subspaces spanned by $|\psi(\theta^\mu)\rangle$ and $|\nabla_\mu\psi\rangle$.

\subsubsection{General two-parameter cases}
For state $|\psi(\boldsymbol{\theta})\rangle$ of the three-level system with two parameters $\boldsymbol{\theta}=(\theta^1,\theta^2)$, the two ``optimal directions" $|\nabla_1\psi\rangle$ and $|\nabla_2\psi\rangle$ are non-orthogonal generally. They interfere with each other in the projective measurement. The attainable DDS is thus decreased. Specifically, for a given state $|\psi(\boldsymbol{\theta}_0)\rangle$, we have the following property:
\emph{the maximal DDS attained by the vertex measurements (MvDDS) equals the square root of the quantum geometric tensor's determinant, i.e., 
\begin{equation}
\max_{\left\{\boldsymbol{E}^v\right\}}[|\boldsymbol{g}^I|]= | \boldsymbol{\mathcal{Q}}|=|\boldsymbol{g}^F|-\mathcal{B}_{12}^{2}/4,\label{eq:bound}
\end{equation}
where the maximization is done over the sets of vertex measurements $\{\hat{\boldsymbol{E}}^v(\boldsymbol{\theta})\}$ with $\boldsymbol{\theta}$ approaches $\boldsymbol{\theta}_0$. It also indicates the square gap between IDQS and MvDDS, i.e., the unattainable square density of  quantum states, is proportional to the square of Berry curvature.} Next, we will prove Eq.~(\ref{eq:bound}) in general cases, then exemplify it with the SU(3) parameterization in \ref{su3}.

\begin{proof}
We prove this property with its equivalent proposition:  the maximal DDS acquired by vertex measurements $\{\hat{\boldsymbol{E}}^v(\boldsymbol{\theta})\}$ in the neighborhood of state $|\psi(\boldsymbol{\theta}+\delta\boldsymbol{\theta})\rangle$ converges to $\sqrt{|\boldsymbol{\mathcal{Q}}(\boldsymbol{\theta})|}$ with $\delta\boldsymbol{\theta}$ approaches zero.  
Specifically, we fix parameters $\boldsymbol{\theta}$ of the vertex measurement $\hat{\boldsymbol{E}}^v(\boldsymbol{\theta})$, then substitute $\boldsymbol{\theta}_0$ with $\boldsymbol{\theta}+\delta\boldsymbol{\theta}$ to study the DDS acquired in the neighborhood of state $|\psi(\boldsymbol{\theta}+\delta\boldsymbol{\theta})\rangle$. The maximization should be done over both of the sets of vertex measurements $\{\hat{\boldsymbol{E}}^v(\boldsymbol{\theta})\}$, i.e., $\{|\Upsilon_1\rangle,|\Upsilon_2\rangle\}$, and the mismatches $\delta \boldsymbol{\theta}$.

We begin with the assumption that the mismatches $\delta\boldsymbol{\theta}$ are small enough to validate the linear approximation
\begin{equation}
|\psi(\boldsymbol{\theta}+\delta\boldsymbol{\theta})\rangle\approx |\psi(\boldsymbol{\theta}+\delta\boldsymbol{\theta})\rangle_1\equiv c_0 |\psi(\boldsymbol{\theta})\rangle+\delta\theta^{\mu}\hat{\nabla}_{\mu}|\psi(\boldsymbol{\theta})\rangle,\label{ps}
\end{equation}
with $c_0\in \mathbb{C}$, $\mu=1,2$. The overlap of the two derivatives is denoted as $\langle\nabla_{1}\psi|\nabla_{2}\psi\rangle=\cos\alpha e^{\mathrm{i}\beta}$, $0\leqslant\alpha\leqslant \pi/2$, $0\leqslant\beta<2\pi$.
The corresponding quantum geometric tensor is
\begin{equation}
\boldsymbol{\mathcal{Q}}(\boldsymbol{\theta})=\left[\begin{array}{cc}
\lambda_{1}\lambda_{1} &\lambda_{1}\lambda_{2} \cos\alpha e^{\mathrm{i}\beta}\\
\lambda_{1}\lambda_{2} \cos\alpha e^{-\mathrm{i}\beta} & \lambda_{2}\lambda_{2}
\end{array}\right],
\end{equation}
with the determinant $|\boldsymbol{\mathcal{Q}}(\boldsymbol{\theta})|=\lambda_1^2\lambda_2^2\sin^2\alpha$.

Firstly, we will prove 
\begin{equation}
|\boldsymbol{g}^I(\boldsymbol{\theta}+\delta\boldsymbol{\theta})|_{\delta\boldsymbol{\theta}\rightarrow\boldsymbol{0}}\leqslant|\boldsymbol{\mathcal{Q}}(\boldsymbol{\theta})|,\label{eq:1}
\end{equation}
by introducing polar parameters $\boldsymbol{\eta}=( r,\theta_{\chi})$ with
\begin{eqnarray}
\delta\theta^{1}\lambda_1 =r\cos\theta_{\chi},&\quad&
\delta\theta^{2}\lambda_2 =r\sin\theta_{\chi},
\end{eqnarray}
$r\geqslant0$, and $0\leqslant\theta_\chi<2\pi$. 
In basis of the vertex measurement $\hat{\boldsymbol{E}}^v(\boldsymbol{\theta})$, we  have
\begin{eqnarray}
|\psi (\boldsymbol{\theta}+\delta\boldsymbol{\theta})\rangle_1 = c_0|\psi(\boldsymbol{\theta})\rangle+\sum_i x_{i}e^{{\rm i}\phi_{i}}|\Upsilon_{i}\rangle,
\end{eqnarray}
with
$
 x_{i}e^{{\rm i}\phi_{i}}=r\left[\cos\theta_{\chi}\langle \Upsilon_{i}|\nabla_{1}\psi\rangle+\sin\theta_{\chi}\langle \Upsilon_{i}|\nabla_{2}\psi\rangle\right],
$
$ x_i\geqslant 0$, and $0\leqslant\phi_i<2$, as functions of $\boldsymbol{\eta}$. For the parameter $\eta^\mu$, we define an alternative derivative
\begin{align}
\tilde{\nabla}_{\mu}|\psi(\boldsymbol{\theta})\rangle&\equiv(\hat{\mathbbm{1}}-|\psi(\boldsymbol{\theta})\rangle\langle\psi(\boldsymbol{\theta})|)\tilde{\partial}_{\mu}|\psi(\boldsymbol{\theta}+\delta\boldsymbol{\theta})\rangle_1\nonumber\\
&= \sum_{i}(\tilde{\partial}_{\mu}x_{i}+\mathrm{i}\tilde{\partial}_{\mu}\phi_{i}x_{i})e^{i\phi_{i}}|\Upsilon_{i}\rangle,
\end{align}
with $\tilde{\partial}_\mu\equiv\partial/\partial\eta^\mu$.
Obviously, the two kinds of derivatives are connected with a Jacobian $\boldsymbol{J}\equiv(\partial \boldsymbol{\eta}/\partial\boldsymbol{\theta})$ as
\begin{equation}
\left[\begin{array}{cc}
\nabla_{1}\\ \nabla_{2}\\
\end{array}\right]=\boldsymbol{J}^T\left[\begin{array}{c}
\tilde{\nabla}_{1}\\ \tilde{\nabla}_{2}\\
\end{array}\right].
\end{equation}
For the parameter $\eta^{1}=r$ only relates to the modes $\{x_i\}$, 
we have $\tilde{\partial}_{1}\phi_{i}=0,$ with $i=1,2$.
Hence, the corresponding ``quantum geometric tensor'' can be simplified as
\begin{align}
\boldsymbol{\tilde{\mathcal{Q}}}= & \left[\begin{array}{cc}
\tilde{g}_{11} & \tilde{g}_{12}+\mathrm{i}\sum_{i}\tilde{\partial}_{1}x_{i}\tilde{\partial}_{2}\phi_{i}x_{i}\\
\tilde{g}_{12}-\mathrm{i}\sum_{i}\tilde{\partial}_{1}x_{i}\tilde{\partial}_{2}\phi_{i}x_{i} & \tilde{g}_{22}+\sum_{i}\tilde{\partial}_{2}\phi_{i}\tilde{\partial}_{2}\phi_{i}x_{i}^{2}
\end{array}\right],\label{Q3}
\end{align}
with $\tilde{\mathcal{Q}}_{\mu\nu}\equiv\langle\psi(\boldsymbol{\theta})|\overleftarrow{\tilde{\nabla}}_\mu\tilde{\nabla}_\nu|\psi(\boldsymbol{\theta})\rangle$ and
\begin{equation}
\tilde{g}_{\mu\nu}\equiv\sum_{i=1,2}\tilde{\partial}_{\mu}x_{i}\tilde{\partial}_{\nu}x_{i}.
\end{equation}
Based on Eq.~(\ref{Q3}) , we have the difference
\begin{align}
|\tilde{\boldsymbol{\mathcal{Q}}}|-|\tilde{\boldsymbol{g}}|=& \tilde{g}_{11}\sum_{i}\tilde{\partial}_{2}\phi_{i}\tilde{\partial}_{2}\phi_{i}x_{i}^{2}- (\sum_{i}\tilde{\partial}_{1}x_{i}\tilde{\partial}_{2}\phi_{i}x_{i})^{2}\nonumber\\
\geqslant& 0,\label{eq:2}
\end{align}
where the equality is reached by $|\psi(\boldsymbol{\theta}+\delta\boldsymbol{\theta})\rangle_1$ with
$
\tilde{\partial}_{1}x_{1}/\tilde{\partial}_{1}x_{2}=\tilde{\partial}_{2}\phi_{1}x_{1}/(\tilde{\partial}_{2}\phi_{2}x_{2}),
$
i.e.,
\begin{equation}
\tilde{\partial}_2(\phi_1-\phi_2)=0,\label{OpCon}
\end{equation}
together with $\tilde{\partial}_1(\phi_1-\phi_2)=0$, this condition indicates the relative phase $(\phi_1-\phi_2)$ is constant in the neighborhood of state $|\psi(\boldsymbol{\theta}+\delta\boldsymbol{\theta})\rangle$. 

With the mismatches $\delta\boldsymbol{\theta}\rightarrow 0$, $\tilde{g}_{\mu\nu}$ converges to the element of FRM $\tilde{g^I}_{\mu\nu}(\boldsymbol{\theta}+\delta\boldsymbol{\theta})$, 
where the term $\tilde{\partial}_\mu p_0\tilde\partial_\nu p_0/p_0$ with $p_0=|c_0|^2$ is null, for $\partial_\mu p_0$ is first-order infinitesimal and $p_0\rightarrow1$. 
Then pre-multiplying $\boldsymbol{J}^T$ and post-multiplying $\boldsymbol{J}$ on both sides of the Eq.~(\ref{eq:2}), we have $\boldsymbol{J}^T\tilde{\boldsymbol{\mathcal{Q}}}\boldsymbol{J}=\boldsymbol{\mathcal{Q}}(\boldsymbol{\theta})$ and $\boldsymbol{J}^T\tilde{\boldsymbol{g}^I}(\boldsymbol{\theta}+\delta\boldsymbol{\theta})\boldsymbol{J}=\boldsymbol{g}^I(\boldsymbol{\theta}+\delta\boldsymbol{\theta})$. The inequality Eq.~(\ref{eq:1}) is thus proved. 

Next, we will show the attainability of Eq.~(\ref{eq:1}) via a specific measurement
\begin{eqnarray}
\hat{\boldsymbol{E}}^1(\boldsymbol{\theta})=\{|\psi(\boldsymbol{\theta})\rangle\langle\psi(\boldsymbol{\theta})|,|\nabla_1\psi\rangle\langle\nabla_1\psi|,|\Upsilon_2\rangle\langle \Upsilon_2|\},
\end{eqnarray}
with
$
|\Upsilon_2\rangle=(|\nabla_2\psi\rangle-\cos\alpha e^{i\beta}|\nabla_1\psi\rangle)/\sin\alpha
$. 
In this basis, we have the coefficients
\begin{eqnarray}
x_1e^{\mathrm{i}\phi_1}&=&r\left[\cos\theta_\chi+\sin\theta_\chi \cos\alpha e^{i\beta}\right],\nonumber\\
x_2e^{\mathrm{i}\phi_2}&=&r\sin\theta_\chi\sin\alpha.
\end{eqnarray}
The condition Eq.~(\ref{OpCon}) is satisfied by states $|\psi(\boldsymbol{\theta}+\delta\boldsymbol{\theta})\rangle$ with $\theta_\chi=0$, i.e., $|\delta\theta^2\lambda_2|/|\delta\theta^1\lambda_2|=0$.  The corresponding FRM with respect to the parameters $(\theta^1,\theta^2)$ is 
\begin{align}
&\left.\boldsymbol{g}^I(\boldsymbol{\theta}+d\boldsymbol{\theta})\right|_{\delta\boldsymbol{\theta}\rightarrow\boldsymbol{0}}\nonumber\\
=&
\left[\begin{array}{cc}
\lambda_1^2& \lambda_1\lambda_2\cos\alpha\cos\beta\\
\lambda_1\lambda_2\cos\alpha\cos\beta&\lambda_2^2(\sin^2\alpha+\cos^2\alpha\cos^2\beta)
\end{array}\right],
\end{align}
and the determinant $|\boldsymbol{g}^I(\boldsymbol{\theta}+d\boldsymbol{\theta})|\rightarrow|\boldsymbol{\mathcal{Q}}(\boldsymbol{\theta})|=\lambda_1^2\lambda_2^2\sin^2\alpha$, with ${\delta\boldsymbol{\theta}\rightarrow\boldsymbol{0}}$. It indicates the measurement $\hat{\boldsymbol{E}}^1(\boldsymbol{\theta})$  asymptotically attains the McDDS for states in the neighborhood of $|\psi(\boldsymbol{\theta}+\delta\boldsymbol{\theta})\rangle$, with $|\delta\theta^2\lambda_2|/|\delta\theta^1\lambda_1|\approx0$ and $|\delta\theta^1\lambda_1|\rightarrow 0$. Together with the inequality Eq. (\ref{eq:1}), we have thus proved the Eq.~(\ref{eq:bound}).
\end{proof}
\begin{figure}[htb]
\centering
\includegraphics[width=8.5cm]{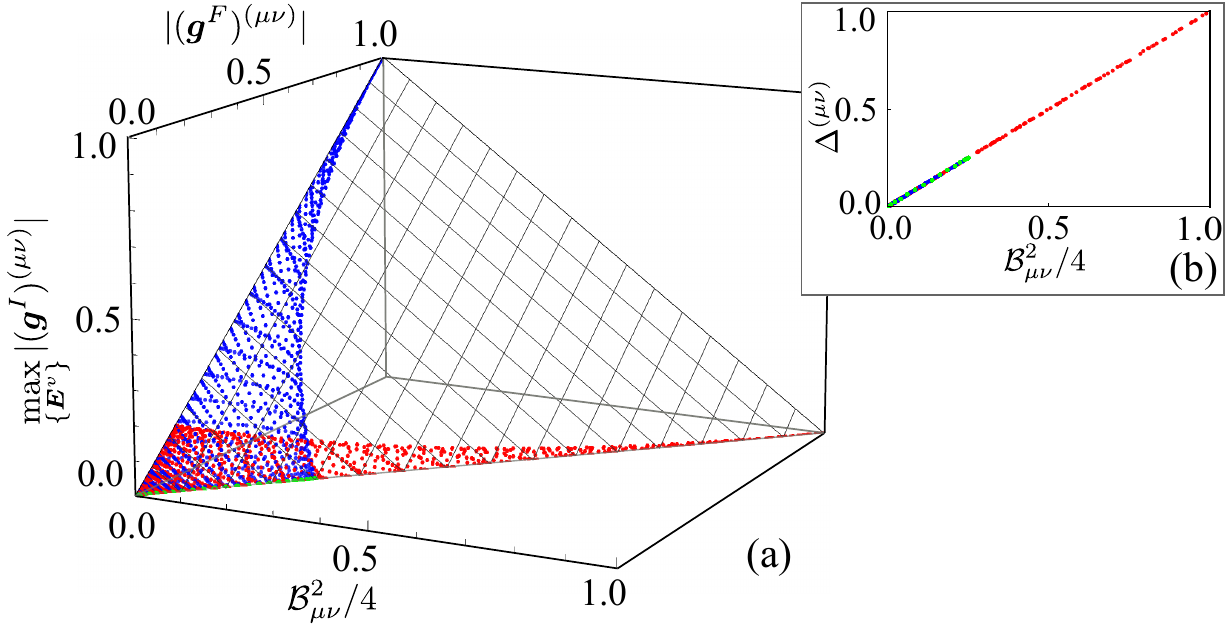}\caption{(color online). Co-distribution of IDQS ($|(\boldsymbol{g}^F)^{(\mu\nu)}|$), MvDDS ($\max_{\{\boldsymbol{E}^v\}}[|(\boldsymbol{g}^I)^{(\mu\nu)}|]$), and Berry curvature ($\mathcal{B}_{\mu\nu}$). The MvDDS is acquired via numerical optimization. (a) The red, blue, and green dots denote states in the sub-manifold $\Theta^{(\alpha\beta)}$, $\Theta^{(\alpha\theta)}$, and $\Theta^{(\gamma\theta)}$, respectively. The plane is given by Eq.~(\ref{eq:bound}). (b) The gap between IDQS and MvDDS ($\Delta^{(\mu\nu)}$) v.s. Berry curvature ($\mathcal{B}_{\mu\nu}$). The gap is defined with  $\Delta^{(\mu\nu)}=|(\boldsymbol{g}^F)^{(\mu\nu)}|-$ $\max_{\{\hat{\boldsymbol{E}}^v\}}[|(\boldsymbol{g}^I)^{(\mu\nu)}|]$. } 
\label{Fig4}
\end{figure}

\subsubsection{SU(3) parameterization\label{su3}}
We parameterize the three-level system with
\begin{eqnarray}
|\psi(\boldsymbol{\theta})\rangle&= & e^{\mathrm{i}(\alpha+\gamma)}\cos\beta\sin\theta|1\rangle-e^{-\mathrm{i}(\alpha-\gamma)}\sin\beta\sin\theta|2\rangle\nonumber \\
 & &+\cos\theta|3\rangle,
\end{eqnarray}
where the parameters $\boldsymbol{\theta}=(\alpha,\gamma,\beta,\theta)$, $0\leqslant \alpha,\gamma < \pi$ and $0\leqslant\beta,\theta<\pi/2$, are the Euler coordinates of the SU(3) group \cite{Herman1966,Byrd1998} with a global phase removed. Via the detailed calculation, we find the quantum geometric tensor of $|\psi(\boldsymbol{\theta})\rangle$ is 
\begin{widetext}
\begin{equation}
\boldsymbol{\mathcal{Q}}^{(\boldsymbol{\theta})}= \left[\begin{array}{cccc}
\sin^{2}\theta\left(1-\cos^{2}2\beta\sin^{2}\theta\right) & \cos2\beta\sin^{2}\theta\cos^{2}\theta & \mathrm{i}\sin2\beta\sin^{2}\theta & -\mathrm{i}\cos2\beta\sin\theta\cos\theta\\
\cos2\beta\sin^{2}\theta\cos^{2}\theta & \sin^{2}\theta\cos^{2}\theta & 0 & -\mathrm{i}\sin\theta\cos\theta\\
-\mathrm{i}\sin2\beta\sin\theta^{2} & 0 & \sin^{2}\theta & 0\\
\mathrm{i}\cos2\beta\sin\theta\cos\theta & \mathrm{i}\sin\theta\cos\theta & 0 & 1
\end{array}\right].
\end{equation}
\end{widetext}
Under this parameterization, we have three two-dimensional sub-manifolds: $\Theta^{(\alpha\beta)}$, $\Theta^{(\alpha\theta)}$, and
$\Theta^{(\gamma\theta)}$, with non-zero Berry curvature. 
We calculate the MvDDS of quantum states in each of them numerically by optimizing the DDS of the sample states over the set of ternary outcome vertex measurements $\hat{\boldsymbol{E}}^v$. With the numerical results, we draw the co-distribution of the IDQS, Berry curvature, and MvDDS in Fig.~\ref{Fig4}. Each of the data points denotes a sample quantum state. All of the data points are located on the plane given by Eq.~(\ref{eq:bound}).  It indicates the square of MvDDS equals determinant of the quantum geometric tensor, i.e., the unattainable square density of  quantum states proportional to the square of Berry curvature. 

\section{Conclusions}
In this article, we have studied the multi-parameter estimation from the information geometry perspective. By taking CFI as (four times of) the metric $\boldsymbol{g}^I$ of the statistical manifold, we proposed a measure $\sqrt{|\boldsymbol{g}^I|}$ named as the density of distinguishable states (DDS). The DDS measures the maximal density of states (estimators) distinguishable in the neighborhood of the $d$-dimensional  estimand $\boldsymbol{\theta}$. The volume of corresponding estimators $\boldsymbol{\theta}_{\rm est}$  depicts the uncertainty of multi-parameter estimation and measured by $\sqrt{|4\boldsymbol{\Sigma}|}$, with $\boldsymbol{\Sigma}$ denoting its covariance matrix. As the  quantum counterpart of CFI, the QFI is four times of the FRM $\boldsymbol{g}^F$, which is the metric of projective Hilbert spaces. The invariant volume elements of $\boldsymbol{g}^F$ defines the intrinsic density of quantum states (IDQS) of the projective Hilbert space with $\sqrt{|\boldsymbol{g}^F|}$.  As an application, IDQS provides us a new method of  calculating the (over) completeness relation of a class of states.  The examples of coherent states and squeezed states have been given. We have proposed a determinant form quantum Cramer-Rao inequality to study the ability to infer the IDQS via quantum measurement and estimation. As a result, we have found that the IDQS bounds the DDS from above. However, different from the single parameter cases, this bound is not  exact generally. Exemplified with the three-level system with two parameters,  we found the square gap between the IDQS and the maximal DDS acquired by vertex measurement equals the square of Berry curvature. It reveals the inner connections between  the gap and the uncertainty principle of quantum theory.
  
Quantifying the distinguishability of quantum states is one of the central topics in studying the statistical aspects of quantum theory. The QFI and statistical distance perform well in single parameter cases. As a qualified measure of the distinguishability, IDQS (DDS) is an essential extension of the statistical distance  in multi-parameter cases. Their values are promising, for ample topics we interested in are generally multi-parameter. Theoretically, the complex projective Hilbert spaces \cite{Bengtsson2006,Anandan1990,Brody2001}, which depicts the fundamental geometrical structures of quantum theory, are intrinsically multi-dimensional. In practical studies such as the ground state manifolds \cite{Kolodrubetz2013}, quantum phase transition \cite{Kumar2014}, response theory \cite{Kolodrubetz2017,Shitara2016,Ozawa2018}, even thermodynamics \cite{Sivak2012,Ruppeiner1979,Ruppeiner1995,Brody1995,Carollo2019}, the systems under investigation are generally multi-dimensional too. By quantifying the distinguishability of quantum states in those cases, the IDQS (DDS) may give impetus to the corresponding studies. 

Precisely, the IDQS also measures the quantum state's overall sensitivity to the small shift of a set of given parameters (both of the intrinsic and external control parameters \cite{You2007,Yang2008,Gu2009,Garnerone2009,Gu2010}). Hence the applications of IDQS to the studies such as the quantum phase transition \cite{Venuti2007,Zanardi2007}, dynamics of open quantum system \cite{Lu2010} are promising. 
The DDS is also an essential measure of multi-dimensional manifolds in classical information geometry. It is potentially a powerful tool to study neural networks, classical statistics, and  thermodynamics.
Furthermore, we have shown that the gap between IDQS and maximal DDS is the signature of the uncertainty principle in the framework of information geometry. It confirms the insights that quantum multi-parameter estimation is a perfect scenario to study the limits of quantum measurements. We wish the further studies may reveal more internal connections between the multi-parameter estimation and quantum measurements.  

\begin{acknowledgments}
HJX thanks Yimu Du for the helpful discussion. This work is supported by the National Natural Science Foundation of China (Grant No. 11725417, Grant No. U1930403), and Science Challenge Project (Grant No. TZ2018005).
\end{acknowledgments}

\appendix

\section{\label{ProofCompleteness}Proof of the completeness relation Eq.~(\ref{CR1})}

We study the $(n+1)$-level system with a set of orthogonal complete basis $\{|0\rangle,|a\rangle|a=1,2,\dots,n\}$. It spans a Hilbert space $\mathcal{H}$ with the completeness relation
\begin{equation}
|0\rangle\langle 0|+\sum_{a}|a\rangle\langle a|=\hat{\mathbbm{1}}.
\end{equation}
 An arbitrary pure state in $\mathcal{H}$ can be expanded as
\begin{equation}
|\psi(\boldsymbol{\theta})\rangle =\sum_{a=1}^{n}x_ae^{{\rm i}\phi_a}|a\rangle+x_0e^{\mathrm{i}\phi_0}|0\rangle,\label{eq:initialstate}
\end{equation}
with $x_0, x_a\geqslant 0$, and the phases $0\leqslant \phi_0,\phi_a<2\pi$. By further introducing the normality
\begin{equation}
x_0^2+\sum_{a}x_a^2=1,
\end{equation}
and fixing the phase $\phi_0=0$,  $|\psi(\boldsymbol{\theta})\rangle$ denotes a quantum state in $\mathbb{C}\boldsymbol{\rm P}^n$ with the real coordinates
\begin{equation}
\boldsymbol{\theta}=(x_1,x_2,\dots,x_n;\phi_1,\phi_2,\dots,\phi_n).
\end{equation}
Based on the above setups, we will show that the IDQS is the measure to construct the identity (completeness relation) of $\mathcal{H}$, i.e.,
\begin{equation}
\hat{\mathbbm{1}}\propto\int d^{2n}\boldsymbol{\theta}\sqrt{|\boldsymbol{g}^F(\boldsymbol{\theta})}|\psi(\boldsymbol{\theta})\rangle\langle\psi(\boldsymbol{\theta})|\equiv\hat{\mathbb{K}}.\label{eq:identity}
\end{equation}

We start from the derivative
\begin{equation}
|d\psi(\boldsymbol{\theta})\rangle=\sum_a\partial_a|\psi(\boldsymbol{\theta})\rangle dx_a+\sum_a\partial'_a|\psi(\boldsymbol{\theta})\rangle d\phi_a,
\end{equation} 
with $\partial_a\equiv\partial/\partial x_a$, $\partial'_a\equiv\partial/\partial\phi_a$, and
\begin{eqnarray}
\partial_a|\psi(\boldsymbol{\theta})\rangle&=&-x_a/x_0|0\rangle+e^{{\rm i}\phi_a}|a\rangle\\
\partial'_a|\psi(\boldsymbol{\theta})\rangle&=&{\rm i}x_a e^{\mathrm{i}\phi_a}|a\rangle.
\end{eqnarray}
Based on it, we have the line element
\begin{eqnarray}
ds^{2}&=&\langle d\psi(\boldsymbol{\theta})|d\psi(\boldsymbol{\theta})\rangle-\langle d\psi(\boldsymbol{\theta})|\psi(\boldsymbol{\theta})\rangle\langle\psi(\boldsymbol{\theta})|d\psi(\boldsymbol{\theta})\rangle\nonumber\\%
&=&\sum_{ab}(\frac{x_{a}x_{b}}{x_{0}^{2}}+\delta_{ab})dx_{a}dx_{b}\nonumber\\
&&+\sum_{ab}(\delta_{ab}x_{a}^{2}-x_{a}^{2}x_{b}^{2})d\phi_{a}d\phi_{b}.\label{Lineelement}
\end{eqnarray}
It indicates the metric
\begin{equation}
\boldsymbol{g}^F=\left[\begin{array}{cc}
(\boldsymbol{g}^F)^{\boldsymbol{x}} & \boldsymbol{0}\\
\boldsymbol{0} & (\boldsymbol{g}^F)^{\boldsymbol{\phi}}
\end{array}\right],
\end{equation}
with
\begin{equation}
(\boldsymbol{g}^F)^{\boldsymbol{x}}=\frac{1}{x_{0}^{2}}\left[\begin{array}{cccc}
x_{1}^{2}+x_{0}^{2} & x_{1}x_{2} & \cdots & x_{n}x_{1}\\
x_{1}x_{2} & x_{2}^{2}+x_{0}^{2} &  & x_{n}x_{2}\\
\vdots &  & \ddots & \vdots\\
x_{1}x_{n} & x_{2}x_{n} & \cdots & x_{n}^{2}+x_{0}^{2}
\end{array}\right],
\end{equation}
\begin{equation}
(\boldsymbol{g}^F)^{\boldsymbol{\phi}}=\left[\begin{array}{cccc}
x_{1}^{2}-x_{1}^{4} & -x_{1}^{2}x_{2}^{2} & \cdots & -x_{n}^{2}x_{1}^{2}\\
-x_{1}^{2}x_{2}^{2} & x_{2}^{2}-x_{2}^{4} &  & -x_{n}^{2}x_{2}^{2}\\
\vdots &  & \ddots & \vdots\\
-x_{1}^{2}x_{n}^{2} & -x_{2}^{2}x_{n}^{2} & \cdots & x_{n}^{2}-x_{n}^{4}
\end{array}\right].
\end{equation}
Hence, we have the IDQS
\begin{equation}
\sqrt{|\boldsymbol{g}^F|}=\prod_a x_a.
\end{equation}
Insert it into the right-hand side of Eq. (\ref{eq:identity}), we have the element
\begin{eqnarray}
\langle c|\hat{\mathbb{K}}|b\rangle&=&\int  d^{n}\boldsymbol{x}\sqrt{|\boldsymbol{g}^F|} x_c x_b\int d^n\boldsymbol{\phi} e^{\mathrm{i}(\phi_b-\phi_c)}\nonumber\\
&=&\mathrm{Vol}(\mathbb{C}\boldsymbol{\mathrm{P}}^n)/(n+1) \delta_{bc},\\
\langle 0|\hat{\mathbb{K}}|0\rangle&=&\int  d^{n}\boldsymbol{x}\sqrt{|\boldsymbol{g}^F|} x_0^2\int d^n\boldsymbol{\phi}\nonumber\\
&=&\mathrm{Vol}(\mathbb{C}\boldsymbol{\mathrm{P}}^n)/(n+1),\\
\langle 0|\hat{\mathbb{K}}|b\rangle&=&\int  d^{n}\boldsymbol{x}\sqrt{|\boldsymbol{g}^F|} x_0 x_b\int d^n\boldsymbol{\phi} e^{\mathrm{i}\phi_b}\nonumber\\
&=&0,
\end{eqnarray}
with the volume of  $\mathbb{C}\boldsymbol{\rm P}^n$
\begin{equation}
{\rm Vol}(\mathbb{C}\boldsymbol{\rm P}^n)\equiv\int d^{2n}\boldsymbol{\theta} \sqrt{|\boldsymbol{g}^F(\boldsymbol{\theta})|}=\frac{\pi^n}{n!}.
\end{equation}
We have thus proved the completeness relation
\begin{equation}
\frac{\pi^n}{(n+1)!}\int d^{2n}\boldsymbol{\boldsymbol{\theta}}\sqrt{|\boldsymbol{g}^F|}|\psi(\boldsymbol{\theta})\rangle\langle\psi(\boldsymbol{\theta})|=\hat{\mathbbm{1}}.
\end{equation}
Furthermore, we mention that the form of this identity is invariant under re-parameterization, hence its validity is independent of the choice of the coordinates.

\section{\label{App:ECR} Examples of calculating the complete relationship with QFM}
\subsection{Coherent states}
The coherent states widely used in the quantum optics and quantum information fields are defined as
\begin{equation}
|\alpha\rangle=e^{\alpha \hat{a}^{\dagger}-\alpha^*\hat{a}}|0\rangle=e^{-|\alpha|^2/2}\sum_{n=0}^{\infty}\frac{\alpha^n}{\sqrt{n!}}|n\rangle,
\end{equation}
where $\hat{a}$ ($\hat{a}^\dagger$)  denotes a boson annihilation (creation) operator, $|n\rangle=\hat{a}^{\dagger n}|0\rangle/\sqrt{n!}$ is the number state, $\alpha$ is a complex number. At first, we separate $\alpha$ into two real parameters with $\alpha=R+\mathrm{i}I$. The parameter space is $\Theta=\mathbb{R}^2$, and the corresponding derivatives are
\begin{eqnarray}
\partial_{R}|\alpha\rangle =(\hat{a}^{\dagger}-R)|\alpha\rangle, &  & 
\partial_{I}|\alpha\rangle =(\mathrm{i}\hat{a}^{\dagger}-I)|\alpha\rangle.
\end{eqnarray}
We have the elements of quantum geometric tensor
\begin{eqnarray}
\mathcal{Q}_{RI} &=& \mathcal{Q}_{IR}^{*}\nonumber\\
&=&\langle\alpha|\overleftarrow{\partial}_{R}\partial_{I}|\alpha\rangle-\langle\alpha|\overleftarrow{\partial}_{R}|\alpha\rangle\langle\alpha|\partial_{I}|\alpha\rangle\nonumber\\
&=& \mathrm{i},\\
\mathcal{Q}_{RR} &= & \mathcal{Q}_{II}\nonumber\\
&=& \langle\alpha|\overleftarrow{\partial}_{R}\partial_{R}|\alpha\rangle-\langle\alpha|\overleftarrow{\partial}_{R}|\alpha\rangle\langle\alpha|\partial_{R}|\alpha\rangle\nonumber\\
&= & 1.
\end{eqnarray}
It indicates $\boldsymbol{\mathcal{Q}}=\boldsymbol{g}^F+\mathrm{i}\boldsymbol{\sigma}$ with
\begin{eqnarray}
\boldsymbol{g}^F=\left[\begin{array}{cc}
1 & 0\\
0 & 1
\end{array}\right], & &
\boldsymbol{\sigma}=\left[\begin{array}{cc}
0 & 1\\
-1 & 0
\end{array}\right]
\end{eqnarray}
in the coordinates $(R,I)$. Hence we have the integral
\begin{eqnarray}
& & \int_{\mathbb{R}^2}dRdI\sqrt{ |\boldsymbol{g}^F|}|\alpha\rangle\langle\alpha|\nonumber\\
&= & \int_{\mathbb{C}}d^2\alpha e^{-|\alpha|^2}\sum_{n,m=0}^{\infty}\frac{\alpha^n\alpha^{*m}}{\sqrt{n!m!}}|n\rangle\langle m|\nonumber\\
&=& 2\pi \sum_{n=0}^{\infty}  \int_0^\infty d|\alpha| e^{-|\alpha|^2}\frac{|\alpha|^{2n}}{n!}|n\rangle\langle n|\nonumber\\
&=&\pi\sum_{n=0}^{\infty} |n\rangle\langle n|
\end{eqnarray}
It is the over completeness relation of coherent states \cite{Scully1997}. We also mention that the metric $\boldsymbol{g}^F$ is Euclidean, which indicates the manifold composed by coherent states is flat and uniform. It is consistent with the understanding  that  this manifold is formed by the shift of the vacuum states $|0\rangle$ with operator $e^{\alpha\hat{a}^\dagger-\alpha^*\hat{a}}$.  
\subsection{Spin squeezed states}
In this part, we take the spin squeezed states or the SU(1,1) coherent states as an example. The states are defined with the SU(1,1) algebra 
\begin{equation}
[\hat{K}_{1},\hat{K}_{2}]=-\mathrm{i}\hat{K}_{0},[\hat{K}_{0},\hat{K}_{1}]=\mathrm{i}\hat{K}_{0},[\hat{K}_{2},\hat{K}_{0}]=\mathrm{i}\hat{K}_{1},
\end{equation}
with the Casimir operator 
\begin{equation}
\hat{C}=\hat{K}_{0}^{2}-\hat{K}_{1}^{2}-\hat{K}_{2}^{2}=\hat{K}_{0}^{2}-\frac{1}{2}(\hat{K}_{+}\hat{K}_{-}+\hat{K}_{-}\hat{K}_{+}).
\end{equation}
The basis vector $|k,m\rangle$ of the unitary irreducible representation is defined by 
\begin{eqnarray}
\hat{C}|k,m\rangle &= & k(k-1)|k,m\rangle\nonumber \\
\hat{K}_{0}|k,m\rangle &= & (k+m)|k,m\rangle,
\end{eqnarray}
where $k$ is the Bargmann index. The basis vectors $\{|k,m\rangle|m\}$ span the corresponding representation spaces.  The completeness relation of this representation is
\begin{equation}
\hat{\mathbbm{1}}=\sum_{m}|k,m\rangle\langle k,m|.
\end{equation}
For single mode squeezed states, $k$ equals $1/4$ ($3/4$) corresponding to the even (odd) particle number space. For two modes squeezed states, we have $k=(n_0+1)/2$, where $n_0$ denotes the number difference between two modes.
 
The SU(1,1) coherent state is defined as
\begin{equation}
|z,k\rangle=\exp(\zeta \hat{K}_{+}-\zeta^{*}\hat{K}_{-})|k,0\rangle,
\end{equation}
with the complex number $z=\zeta/|\zeta|\tanh|\zeta|$ locates in a disk $D=\{z||z|<1\}$. In the basis $|k,m\rangle$, the SU(1,1) coherent state can be expanded as 
\begin{equation}
|z,k\rangle=(1-|z|^{2})^{k}\sum_{m=0}^{\infty}\sqrt{\frac{\Gamma(2k+m)}{m!\Gamma(2k)}}z^{m}|k,m\rangle.
\end{equation}
Via a detailed calculation, we find the QFM is 
\begin{equation}
\boldsymbol{g}^F=\frac{k}{2(1-|z|^{2})^{2}}\left[\begin{array}{cc}
1\\
 & |z|^{2}
\end{array}\right]
\end{equation}
in the coordinates $(|z|,\theta)$ with $z=|z|e^{\mathrm{i}\theta}$. If $k > 1/2$, we have the integral over the disk $D$ as
\begin{widetext} 
\begin{eqnarray}
 & & \int_D d|z|d\theta\sqrt{|\boldsymbol{g}^F|}|z,k\rangle\langle z,k|\nonumber\\
&= & \sum_{m,n=0}\int_0^1 d|z|\int_0^{2\pi} d\theta\frac{k}{2(1-|z|^{2})^{2}}|z|(1-|z|^{2})^{2k}\sqrt{\frac{\Gamma(2k+m)\Gamma(2k+n)}{m!\Gamma(2k)\Gamma(2k)n!}}z^{m}z^{*n}|k,m\rangle\langle k,n|\nonumber\\
&= & \sum_{m=0}\frac{k\pi\Gamma(2k+m)}{m!\Gamma(2k)}\int_0^1 d|z|\frac{|z|^{2m+1}}{(1-|z|^{2})^{2-2k}}|k,m\rangle\langle k,m|\label{IntUnsolve}\\
&= & \sum_{m=0}\frac{k\pi\Gamma(2k+m)}{2m!\Gamma(2k)}\frac{\Gamma(2k-1)m!}{\Gamma(2k+m)}|k,m\rangle\langle k,m|\nonumber\\
&= & \frac{k\pi}{2(2k-1)}\sum_{m=0}|k,m\rangle\langle k,m|.
\end{eqnarray}

Obviously, this integral is proportional to $\hat{\mathbbm{1}}$. Hence we have the identity 
\begin{equation}
\hat{\mathbbm{1}}^{(k)}=\frac{2(2k-1)}{k\pi}\int_{D}d\Theta\sqrt{|\boldsymbol{g}^F|}|z\rangle\langle z|.
\end{equation}

\end{widetext}


\begin{thebibliography}{99}

\bibitem{Szczykulska2016}M. Szczykulska, T. Baumgratz, and A. Datta, \textit{Multi-parameter Quantum Metrology}, Advances in Physics: X \textbf{1}, 621 (2016).

\bibitem{Liu2020} J. Liu, H. Yuan, X.-M. Lu, and X. G. Wang, \textit{Quantum Fisher Information Matrix and Multiparameter Estimation}, J. Phys. A: Math. Theor. \textbf{53}, 023001 (2020).

\bibitem{Matsumoto2002}K. Matsumoto, \textit{A New Approach to the Cram\'er-Rao-type Bound of the Pure-state Model}, J. Phys. A: Math. Gen. \textbf{35}, 3111 (2002).

\bibitem{Humphreys2013}P. C. Humphreys, M. Barbieri, A. Datta, and I. A. Walmsley, \textit{Quantum Enhanced Multiple Phase Estimation}, Phys. Rev. Lett. \textbf{111}, 070403 (2013).

\bibitem{Vaneph2013}C. Vaneph, T. Tufarelli, and M. G. Genoni, \textit{Quantum Estimation of a Two-Phase Spin Rotation}, Quantum Meas. Quantum Metro.  \textbf{1}. 12 (2013).

\bibitem{Ragy2016}S. Ragy, M. Jarzyna, and R. Demkowicz-Dobrz\'anski, \textit{Compatibility in Multiparameter Quantum Metrology}, Phys. Rev. A \textbf{94}, 052108 (2016).

\bibitem{Baumgrat2016}T. Baumgratz and A. Datta, \textit{Quantum Enhanced Estimation of a Multidimensional Field}, Phys. Rev. Lett. \textbf{116}, 030801 (2016).

\bibitem{Pezze2017}L. Pezz\'e, M. A. Ciampini,
N. Spagnolo, P. C. Humphreys, A. Datta, I. A. Walmsley, M. Barbieri,
F. Sciarrino, and A. Smerzi, \textit{Optimal Measurements for Simultaneous Quantum Estimation of Multiple Phases}, Phys. Rev. Lett. \textbf{119}, 130504 (2017).

\bibitem{Gessner2018}M. Gessner, L. Pezz\'e, and A. Smerzi, \textit{Sensitivity Bounds for Multiparameter Quantum Metrology}, Phys. Rev. Lett. \textbf{121}, 130503 (2018).

\bibitem{Sidhu2019}J. S. Sidhu, Y. Ouyang, E, T. Campbell, and P. Kok, \textit{Tight Bounds on the Simultaneous Estimation of Incompatible Parameter}, arXiv: 1912.09218v1 (2019).

\bibitem{Vidrighin2014}M. D. Vidrighin, G. Donati, M. G. Genoni, X.-M. Jin, W. S. Kolthammer, M.S. Kim, A. Datta, M. Barbieri, and I. A. Walmsley, \textit{Joint Estimation of Phase and Phase Diffusion for Quantum Metrology}, Nat. Commu. \textbf{5}, 3532 (2014).

\bibitem{Crowley2014}P. J. D. Crowley, A. Datta, M. Barbieri, and I. A. Walmsley, \textit{Tradeoff in Simultaneous Quantum-Limited Phase and Loss Estimation in Interferometry}, Phys. Rev. A \textbf{89}, 023845 (2014).

\bibitem{Kok2017}P. Kok, J. Dunningham, and J. F. Ralph, \textit{Role of Entanglement in Calibrating Optical Quantum Gyroscopes}, Phys. Rev. A \textbf{95}, 012326 (2017).

\bibitem{Helstrom1976} C. W. Helstrom, \textit{Quantum Detection and Estimation Theory} (Academic, New York, 1976).

\bibitem{Holevo1982}A. S. Holevo, \textit{Probabilistic and Statistical Aspect of Quantum Theory} (North-Holland, Amsterdam, 1982).

\bibitem{GLM2004}V. Giovannetti, S. Lloyd, and L. Maccone, \textit{Quantum-Enhanced Measurements: Beating the Standard Quantum Limit}, Science \textbf{306}, 1330 (2004).

\bibitem{GLM2006}V. Giovannetti, S. Lloyd, and L. Maccone, \textit{Quanutm Metrology}, Phys. Rev. Lett. \textbf{96}, 010401 (2006), 

\bibitem{GLM2011}V. Giovannetti, S. Lloyd, and L. Maccone,\textit{Advances in Quantum Metrology}, Nat. Photon. \textbf{5}, 222 (2011).

\bibitem{Pezze2018}L. Pezz\'e, A. Smerzi, M. K. Oberthaler, R. Schmied, and
P. Treutlein, \textit{Quantum Metrology with Nonclassical States of Atomic Ensembles}, Rev. Mod. Phys. \textbf{90}, 035005 (2018), and references therein.

\bibitem{Degen2017}C. L. Degen, F. Reinhard, and P. Cappellaro, \textit{Quantum Sensing}, Rev. Mod. Phys. \textbf{89}, 035002 (2017), and references therein.

\bibitem{Braun2018}D. Braun, G. Adesso, F. Benatti, R. Floreanini, U. Marzolino, M. W. Mitchell, and S. Pirandola, \textit{Quantum-Enhanced Measurements without Entanglement}, Rev. Mod. Phys. \textbf{90}, 035006 (2018), and references therein. 

\bibitem{Swell2012}R. J. Sewell, M. Koschorreck, M. Napolitano, B. Dubost, N. Behbood, and M. W. Mitchell, \textit{Magnetic Sensitivity beyond the Projection Noise Limit by Spin Squeezing}, Phys. Rev. Lett. \textbf{109}, 253605 (2012).

\bibitem{Ockeloen2013} C. F. Ockeloen, R. Schmied, M. F. Riedel, and P. Treutlein, \textit{Quantum Metrology with a Scanning Probe Atom Interferometer}, Phys. Rev. Lett. \textbf{111}, 143001 (2013). 

\bibitem{Muessel2014} W. Muessel, H. Strobel, D. Linnemann, D. B. Hume, and M. K. Oberthaler, \textit{Scalable Spin Squeezing for Quantum-Enhanced Magnetometry with Bose-Einstein Condensates}, Phys. Rev. Lett. \textbf{113}, 103004 (2014).

\bibitem{Louchet-Chauvet2010} A. Louchet-Chauvet, J. Appel, J. J. Renema, D. Oblak, N. Kjrgaard, and E. S. Polzik, \textit{Entanglement-Assisted Atomic Clock Beyond the Projection Noise Limit}, New J. Phys. \textbf{12}, 065032 (2010).

\bibitem{Leroux2010} I. D. Leroux, M. H. Schleier-Smith, and V. Vuleti\'c, \textit{Orientation-Dependent Entanglement Lifetime in a Squeezed Atomic Clock}, Phys. Rev. Lett. \textbf{104}, 250801 (2010).

\bibitem{Hosten2016} O. Hosten, N. J. Engelsen, R. Krishnakumar, and M. A. Kasevich, \textit{Measurement Noise 100 Times Lower than the Quantum-Projection Limit Using Entangled Atoms}, Nature (London) \textbf{529}, 505 (2016).

\bibitem{Kruse2016} I. Kruse, K. Lange, J. Peise, B. L\"ucke, L. Pezz\'e, J. Arlt, W. Ertmer, C. Lisdat, L. Santos, A. Smerzi, and C. Klempt, \textit{Improvement of an Atomic Clock Using Squeezed Vacuum}, Phys. Rev. Lett. \textbf{117}, 143004 (2016).

\bibitem{LIGO2011} The LIGO Scientific Collaboration, \textit{A Gravitational Wave Observatory Operating beyond the Quantum Shot-Noise Limit}, Nat. Phys. \textbf{7}, 962 (2011).

\bibitem{LOGO2013}The LIGO Scientific Collaboration, \textit{Enhanced Sensitivity of the LIGO Gravitational Wave Detector by Using Squeezed States of Lights}, Nat. Photon. \textbf{7}, 613 (2013).

\bibitem{Rao1945}  C. R. Rao, \textit{Information and Accuracy Attainable in the Estimation of Statistical Parameters}, Bull. Calcutta Math. Soc. \textbf{37}, 81(1945).

\bibitem{Amari1985}Shun-ichi Amari, \textit{Differential Geometrical Methods in Statistics}, (Springer-Verlag, Berlin, 1985).

\bibitem{Amari2000}Shun-ichi Amari and H. Nagaoka, \textit{Methods of Information Geometry}, (Oxford university press, New York, 2000).

\bibitem{Amari2016}Shun-ichi Amari, \textit{Information Geometry and Its Applications}, (Springer, Japan, 2016).  

\bibitem{Shapere1989} A. Shapere and F. Wilczek (eds.), \textit{Geometric Phases in Physics} (World Scientific, Singapore, 1989).

\bibitem{Bengtsson2006} I. Bengtsson and K. Zyczkowski, \textit{Geometry of Quantum States} (Cambridge university press, New York, 2006).

\bibitem{Anandan1990} J. Anandan, and Y. Aharonov, \textit{Geometry of quantum evolution}, Phys. Rev. Lett. \textbf{65}, 1697 (1990).

\bibitem{Brody2001} D. C. Brody and L. P. Hughston, \textit{Geometric Quantum Mechanics}, J. Geom. Phys. \textbf{38}, 19 (2001). 

\bibitem{Wootters1981}W. K. Wootters, \textit{Statistical Distance and Hilbert Space}, Phys. Rev. D \textbf{23}, 357 (1981).

\bibitem{Braunstein1994}S. L. Braunstein and C. M. Caves, \textit{Statistical Distance and the Geometry of Quantum States}, Phys. Rev. Lett. \textbf{72}, 3439 (1994).

\bibitem{Gibbons1992} G. W. Gibbons, \textit{Typical States and Density Matrics}, J. Geom. Phys. \textbf{8}, 147 (1992).

\bibitem{You2007} W.-L. You, Y.-W. Li, and S.-J. Gu, \textit{Fidelity, Dynamic Structure Factor, and Susceptibility in Critical Phenomena}, Phys. Rev. E \textbf{76}, 022101 (2007).
\bibitem{Yang2008} S. Yang, S.-J. Gu, C.-P. Sun, and H.-Q. Lin, \textit{Fidelity Susceptibility and Long-Range Correlation in the Kitaev Honeycomb Model}, Phys. Rev.
A \textbf{78}, 012304 (2008).

\bibitem{Gu2009} S.-J. Gu, \textit{Fidelity Susceptibility and Quantum Adiabatic Condition in Thermodynamic Limits}, Phys. Rev. E \textbf{79}, 061125 (2009).

\bibitem{Garnerone2009}S. Garnerone, D. Abasto, S. Haas,
and P. Zanardi, \textit{Fidelity in Topological Quantum Phases of Matter}, Phys. Rev. A \textbf{79}, 032302 (2009).

\bibitem{Gu2010}S.-J. Gu, \textit{Fidelity Approach to Quantum Phase Transitions}, Int. J. Mod. Phys. B \textbf{24}, 4371 (2010).

\bibitem{Venuti2007}L. Campos Venuti and P. Zanardi, \textit{Quantum Critical Scaling of the Geometric Tensors}, Phys. Rev. Lett. \textbf{99}, 095701 (2007). 

\bibitem{Zanardi2007}P. Zanardi, P. Giorda, and M. Cozzini, \textit{Information-Theoretic Differential Geometry of Quantum Phase Transitions}, ,Phys. Rev. Lett. \textbf{99}, 100603 (2007).

\bibitem{Lu2010} X. M. Lu, X. Wang, and C. P. Sun, \textit{Quantum Fisher Information Flow and Non-Markovian Processes of Open Systems}, Phys. Rev. A 82, 042103 (2010).

\bibitem{Jones2010} P. J. Jones and P. Kok, \textit{Geometric Derivation of the Quantum Speed Limit}, Phys. Rev. A \textbf{82}, 022107 (2010). 

\bibitem{Zwierz2012} M. Zwierz, \textit{Comment on ``Geometric Derivation of the Quantum Speed Limit"}, Phys. Rev. A \textbf{86}, 016101 (2012).

\bibitem{Taddei2013}M. M. Taddei, B. M. Escher, L. Davidovich, and R. L. de Matos Filho, \textit{Quantum Speed Limit for Physical Processes}, Phys. Rev. Lett. \textbf{110}, 050402 (2013).

\bibitem{Pires2016} D. P. Pires, M. Cianciaruso, L. C. C\'eleri, G. Adesso, and D. O. Soares-Pinto, \textit{Generalized Geometric Quantum Speed Limits}, Phys. Rev. X \textbf{6}, 021031 (2016).

\bibitem{Bukov2019} M. Bukov, D. Sels, and A. Polkovnikov, \textit{Geometric Speed Limit of Accessible Many-body State Preparation}, Phys. Rev. X \textbf{9}, 011034 (2019). 

\bibitem{Tomka2016} M. Tomka, T. Souza, S. Rosenberg, and A. Polkovnikov, \textit{Geodesic Paths for Quantum Many-Body Systems}, arXiv: 1606. 05890v2 (2016).

\bibitem{Sivak2012} D. A. Sivak and G. E. Crooks, \textit{Thermodynamic Metrics and Optimal Path}, Phys. Rev. Lett. \textbf{108}, 190602 (2012).

\bibitem{Rotskoff2015} G. M. Rotskoff and G. E. Crooks, \textit{Optimal Control in Nonequilibrium Systems: Dynamic Riemannian Geometry of the Ising Model}, Phys. Rev. E \textbf{92}, 060102(R) (2015).

\bibitem{Zulkowski2015} P. R. Zulkowski and M. R. DeWeese, \textit{Optimal Control of Overdamped Systems}, Phys. Rev. E \textbf{92}, 032117 (2015).

\bibitem{Sivak2016}D. A. Sivak and G. E. Crooks, \textit{Thermodynamic Geometry of Minimum-Dissipation Driven Barrier Crossing}, Phys. Rev. E \textbf{94}, 052106 (2016).

\bibitem{Miyake2001} A. Miyake and M. Wadati, \textit{Geometric Strategy for the Optimal Quantum Search}, Phys. Rev. A \textbf{64}, 042317 (2001).

\bibitem{Cafaro2012A} C. Cafaro and S. Mancini, \textit{An Information Geometric Viewpoint of Algorithms in Quantum Computing}, AIP Conf. Proc. \textbf{1443}, 374 (2012).
 
\bibitem{Cafaro2012B}C. Cafaro and S. Mancini, \textit{On Grover's Search Algorithm from a Quantum Information Geometry Viewpoint}, Phys. A \textbf{391}, 1610 (2012).

\bibitem{Amari1992}S.-i. Amari, K. Kurata, and H. Nagaoka, \textit{Information Geometry of Boltzmann Machines}, IEEE, Transactions on Neural networks, \textbf{3}, 260 (1992).

\bibitem{Crooks2007}G. E. Crooks, \textit{Measuring Thermodynamic Length}, Phys. Rev. Lett. \textbf{99}, 100602 (2007).

\bibitem{Zulkowski2012}P. R. Zulkowski, D. A. Sivak, G. E. Crooks, and M. R.
DeWeese, \textit{Geometry of Thermodynamic Control}, Phys. Rev. E 86, 041148 (2012).

\bibitem{Weinhold1975} F. Weinhold, \textit{Metric Geometry of Equilibrium Thermodynamics}, J. Chem. Phys. \textbf{63}, 2479 (1975).

\bibitem{Salamon1980} P. Salamon, A. Nitzan, B. Andresen, and R. S. Berry, \textit{Minimum Entropy Production and the Optimization of Heat Engines}, Phys. Rev. A \textbf{21}, 2115 (1980).

\bibitem{Ruppeiner1979}G. Ruppeiner, \textit{Thermodynamics: A Riemannian Geometric Model},  Phys. Rev. A \textbf{20}, 1608 (1979).

\bibitem{Ruppeiner1995}G. Ruppeiner, \textit{Riemannian Geometry in Thermodynamic Fluctuation Theory}, Rev. Mod. Phys. \textbf{67}, 605 (1995).

\bibitem{Kolodrubetz2013} M. Kolodrubetz, V. Gritsev, and A. Polkovnikov, \textit{Classifying and Measuring Geometry of a Quantum Ground State Manifold}, Phys. Rev. B \textbf{88}, 064304 (2013).

\bibitem{Kumar2014}P. Kumar and T. Sarkar, \textit{Geometric Critical Exponents in Classical and Quantum Phase Transitions}, Phys. Rev. E \textbf{90}, 042145 (2014).

\bibitem{Banchi2014} L. Banchi, P. Giorda, and P. Zanardi, \textit{Quantum Information-Geometry of Dissipative Quantum Phase Transitions}, Phys. Rev. E \textbf{89}, 022102 (2014).

\bibitem{Ozawa2018}T. Ozawa, \textit{Steady-State Hall Response and Quantum Geometry of Driven-Dissipative Lattices}, Phys. Rev. B \textbf{97}, 041108(R) (2018).

\bibitem{Kolodrubetz2017}M. Kolodrubetz, D. Sels, P. Mehta, and A. Polkovnikov, \textit{Geometry and Non-Adiabatic Response in Quantum and Classical Systems}, Phys. Rep. \textbf{697}, 1 (2017).

\bibitem{Shitara2016}T. Shitara and M. Ueda, \textit{Determining the Continuous Family of Quantum Fisher Information from Linear-Response Theory}, Phys. Rev. A \textbf{94}, 062316 (2016).

\bibitem{Brody1995}D. Brody and N. Rivier, \textit{Geometrical Aspects of Statistical Mechanics}, Phys. Rev. E \textbf{51}, 1006 (1995).

\bibitem{Carollo2019} A. Carollo, D. Valenti, D. Spagnolo, \textit{Geometry of Quantum Phase Transitions}, Phys. Rep. \textbf{10}.  1016 (2019).

\bibitem{Lehmann1998} E. L. Lehmann and G. Casella, \textit{Theory of Point Estimation, 2nd edition, Sec. 6.5}. 

\bibitem{JunShao2003}J. Shao, \textit{Mathematical Statistics, 2nd edition, Sec. 4.5} (Springer, New York, 2003).

\bibitem{Jeffreys1946} H. Jeffreys, \textit{An Invariant Form for the Prior Probability in Estimation Problems}, Proc. Roy. Soc. A \textbf{186}, 453 (1946).

\bibitem{Jeffreys1948}H. Jeffereys, \textit{Theory of Probability, 2nd} (Oxford university press, 1948). 

\bibitem{Jaynes1968} E. T. Jaynes, \textit{Prior Probabilities}, IEEE Trans. Syst. Sci. Cybern. \textbf{4}, 227 (1968).

\bibitem{Jaynes2003} E. T. Jaynes,  \textit{Probability Theory: the Logic of Science}, (Cambridge university press, Cambridge, 2003).

\bibitem{Herman1966}R. Hermann, \textit{Lie Groups for Physicists} (Benjamin, New York, 1966).

\bibitem{Byrd1998}M. Byrd, \textit{Differential Geometry on SU(3) with Applications to the Three State Systems}, J. Math. Phys. \textbf{39}, 6125 (1998).

\bibitem{Scully1997} M. O. Scully, M. S. Zubairy, \textit{Quantum Optics} (Cambridge University, 1997).
 
\end{thebibliography}
\end{document}